\documentclass[english,aps,reprint,longbibliography,prx]{revtex4-1}
\usepackage[T1]{fontenc}
\usepackage[utf8x]{inputenc}
\usepackage{color}
\usepackage{babel}
\usepackage{array}
\usepackage{mathtools}
\usepackage{multirow}
\usepackage{amsmath}
\usepackage{amssymb}
\usepackage{graphicx}
\usepackage[pdftex,unicode=true,
 bookmarks=true,bookmarksnumbered=false,bookmarksopen=false,
 breaklinks=false,pdfborder={0 0 1},backref=false,colorlinks=true]
 {hyperref}
\hypersetup{
 pdfauthor={Ying-Jen Yang},
 pdfborderstyle=,allcolors=blue}

\makeatletter

\providecommand{\tabularnewline}{\\}

\usepackage{babel}
\usepackage{cleveref}

\makeatother

\begin{document}
\title{Unified formalism for entropy productions and fluctuation relations}
\author{Ying-Jen Yang}
\email{yangyj@uw.edu}

\affiliation{Department of Applied Mathematics, University of Washington, Seattle,
WA 98195-3925, USA}
\author{Hong Qian}
\email{hqian@uw.edu}

\affiliation{Department of Applied Mathematics, University of Washington, Seattle,
WA 98195-3925, USA}
\begin{abstract}
Stochastic entropy production, which quantifies the difference between
the probabilities of trajectories of a stochastic dynamics and its
time reversals, has a central role in nonequilibrium thermodynamics.
In the theory of probability, the change in the statistical properties
of observables due to reversals can be represented by a change in
the probability measure.\textcolor{blue}{{} }We consider operators on
the space of probability measure that induce changes in the statistical
properties of a process, and formulate entropy productions in terms
of these change-of-probability-measure (CPM) operators.\textcolor{blue}{{}
}This mathematical underpinning of the origin of entropy productions
allows us to achieve an organization of various forms of fluctuation
relations: All entropy productions have a non-negative mean value,
admit the integral fluctuation theorem, and satisfy a rather general
fluctuation relation. Other results such as the transient fluctuation
theorem and detailed fluctuation theorems then are derived from the
general fluctuation relation with more constraints on the operator
of a entropy production.\textcolor{blue}{{}  }We use a discrete-time,
discrete-state-space Markov process to draw the contradistinction
among three reversals of a process: time reversal, protocol reversal
and the dual process. The properties of their corresponding CPM operators
are examined, and the domains of validity of various fluctuation relations
for entropy productions in physics and chemistry are revealed. We
also show that our CPM operator formalism can help us rather easily
extend other fluctuations relations for excess work and heat, discuss
the martingale properties of entropy productions, and derive the stochastic
integral formulas for entropy productions in constant-noise diffusion
process with Girsanov theorem.\textcolor{blue}{{} }Our formalism provides
a general and concise way to study the properties of entropy-related
quantities in stochastic thermodynamics and information theory.
\end{abstract}
\maketitle

\section{Introduction}

Stochastic thermodynamics is a milestone extending equilibrium statistical
physics to the nonequilibrium realm, and could provide a general theory
for emergent phenomena in mesoscopic systems \citep{searles_fluctuation_1999,jiang_mathematical_2004,seifert_stochastic_2012,qian_nonlinear_2016,thompson_nonlinear_2016}.
It studies entropy productions (EPs), their relations to work and
heat done by the system of interest \citep{crooks_nonequilibrium_1998},
 their statistical properties such as expectations and martingale
properties, and also how their probability density functions change
after time reversals, called fluctuation theorems or fluctuation relations
(FRs) \cite{jarzynski_nonequilibrium_1997,crooks_entropy_1999,maes_origin_2004,seifert_stochastic_2012}.

Various distinct FRs for different EPs have been studied in various
settings including \textit{\emph{discrete-time Markov chains \citep{qian_nonequilibrium_2001,crooks_nonequilibrium_1998,crooks_path-ensemble_2000,riechers_fluctuations_2017},
continuous-time Markov chains (Markov jump processes) \citep{lebowitz_gallavotticohen-type_1999,ge_physical_2010,esposito_three_2010,rao_detailed_2018,ge_extended_2009,seifert_entropy_2005},
diffusion processes (as stochastic differential equations or Langevin
equations) \citep{hatano_steady-state_2001,kurchan_fluctuation_1998,qian_mesoscopic_2001,qian_nonequilibrium_2001,ge_extended_2009,seifert_entropy_2005,chernyak_path-integral_2006}}},
and even general stochastic processes \textit{\emph{\cite{crooks_entropy_1999,jiang_mathematical_2004,ge_transient_2007,ge_extended_2009,shargel_measure-theoretic_2010,chetrite_two_2011}}}.
\textcolor{blue}{}\textit{\emph{To discuss a few, in \citep{crooks_entropy_1999},
Crooks' fluctuation theorem for the total entropy production was introduced
for systems with detailed balance and the conditions for it to hold
was illustrated; in \citep{ge_transient_2007}, the dissipation function
from Evans and Searles \cite{evans_fluctuation_2002} was rigorously
shown to generally admit the transient fluctuation theorem (TFT);
in \citep{esposito_three_2010}, a detailed fluctuation theorem related
to the involutive property of the change in probability (iDFT) was
introduced but the correct condition for total entropy production
and non-adiabatic entropy production was not stated; in \citep{seifert_stochastic_2012},
the generalized Crooks' fluctuation theorem for non-detail balanced
systems (rDFT) and TFT for different EPs were discussed thoroughly
for diffusion processes. iDFT was also discussed briefly; and in \citep{garcia-garcia_joint_2012},
rDFT was thoroughly discussed in general Markov processes for the
three entropy productions except dissipation function.}}

The extensiveness of FTs calls for an unifying formalism to derive
all FTs mentioned above in one general theory, organize their domain
of validity comprehensively, and open ways to reveal more properties
of EPs.\textit{\emph{ It is suggested in \cite{jiang_mathematical_2004,ge_transient_2007,ge_extended_2009,wojtkowski_abstract_2009,shargel_measure-theoretic_2010,chetrite_two_2011}
that measure-theoretic probability theory pioneered by A. N. Komogorov
\citep{kolmogorov_foundations_2018} will do the trick. One key is
the understanding that EPs in physics are the }}\textit{fluctuating}\textit{\emph{
}}\textit{relative}\textit{\emph{ }}\textit{entropy }\textit{\emph{for
trajectories between the original process and the reversed process,
with different EPs given by different reversals or composite of reversals
\cite{crooks_nonequilibrium_1998,crooks_entropy_1999,seifert_stochastic_2012}.
In the measure-theoretic formalism, EP is thus mathematically expressed
as the negative logarithm of the Radon-Nikodym Derivative (RND), which
is a re-weighting factor in taking expectation to change a probability
measure from the original one to the other. }}It is this mathematical
underpinning that enables us to arrive an organization of FRs and
further derive/recover other properties with a deeper understanding
of EPs, \emph{e.g.} their martingale properties \cite{chetrite_two_2011,neri_statistics_2017,chetrite_martingale_2019}.
\textcolor{blue}{}

\textcolor{blue}{}Our paper thus serves as a comprehensive overview
on how to understand EPs and their statistical properties, primarily
FRs, from measure-theoretic probability theory and our change-of-probability-measure
(CPM) operator theory. We revisit discussions on different time reversals,
their associated EPs in physics and chemistry, and derive various
FRs from our general and concise approach. Our goal is to demonstrate
that by adopting this CPM operator formalism, one can neatly derive
many known results in the literature with more rigor and generality,
achieve new understanding on EPs and FRs, and reveal more properties
of EPs.

The outline of this paper is summarized below. In Section \ref{sec:General-Theories},
we briefly introduce measure-theoretic probability theory, the notion
of CPM operator, and present general statistical properties for a
general EP as a fluctuating relative entropy in general stochastic
processes \emph{without} Markovian assumption. With more constraints
on the properties of the CPM operator, we deduce the \textit{\emph{general
conditions for various known FRs and reveal a hierarchy of the domain
of validity for various FRs with new relations recognized between
them such as TFT$\Rightarrow$iDFT.}} In Section \ref{sec:Different-Types-of-reversals},
we further use a discrete-time Markov chain to illustrate the contradistinction
of three different time reversals of the dynamics that are prominent
in physics and chemistry. The involutive and commutative properties
of their corresponding CPM operators are discussed.

In Section \ref{sec:Entropy-Productions-in Physics and Chemistry},
we apply the results in the previous two sections to discuss the properties
of the four EPs commonly considered in physics and chemistry. Notably,
we discuss the difference between dissipation function and total entropy
production and show that the two EPs have non-zero difference in expectation
for finite time interval in time homogeneous processes but have the
same entropy production rate in infinitesimal time interval. The two
seemly contradicting results are resolved by noting the non-additivity
of the dissipation function when connecting time intervals.

We further demonstrate how properties of EPs other than their FRs
can be derived and extended in a rather straightforward way with this
CPM operator formalism. We discuss the martingale properties of the
four EPs which can lead to more statistics on the EPs \cite{chetrite_two_2011,neri_statistics_2017,chetrite_martingale_2019},
and extend the so-called differential FR for work and heat \citep{jarzynski_hamiltonian_2000,maragakis_differential_2008}
to non-equilibrium systems at the end of Section \ref{sec:Entropy-Productions-in Physics and Chemistry}.
We also show how to use our CPM operator formalism and Girsanov theorem
\cite{jiang_mathematical_2004} to derive the stochastic integral
formulas of the four EPs for general time inhomogeneous constant-noise
diffusion processes in Section \ref{sec:Entropy-Productions-in Diffusion}.
The notations for the five heavily-discussed FRs and the EPs involved
in them are summarized in Table \ref{tab:The-five-mainly FR}. 
\begin{table*}
\begin{centering}
\begin{tabular}{|c|c|c|c|}
\hline 
FR &
definition &
the EP involved &
validity and sufficient conditions\tabularnewline
\hline 
\hline 
\multirow{2}{*}{GFR} &
\multirow{2}{*}{$\mathbb{P}\left\{ S_{\nu}\in\mathrm{d}s\right\} =e^{s}\mathbb{P}^{\nu}\left\{ S_{\nu}\in\mathrm{d}s\right\} $} &
\multirow{4}{*}{$S_{\nu}(\omega)=\ln\frac{\mathrm{d}\mathbb{P}\ \thinspace}{\mathrm{d}\mathbb{P}^{\nu}}(\omega)$} &
\multirow{4}{*}{generally valid}\tabularnewline
 &  &  & \tabularnewline
\cline{1-2} \cline{2-2} 
\multirow{2}{*}{IFT} &
\multirow{2}{*}{$\mathbb{E}\left[e^{-S_{\nu}}\right]=1$} &  & \tabularnewline
 &  &  & \tabularnewline
\hline 
\multirow{2}{*}{rDFT} &
\multirow{2}{*}{$\mathbb{P}\left\{ S_{\nu}(\omega)\in\mathrm{d}s\right\} =e^{s}\mathbb{P}^{\nu}\left\{ \bar{S}_{\nu}(r(\omega))\in-\mathrm{d}s\right\} $} &
\multirow{2}{*}{$\bar{S}_{\nu}(\omega)=\ln\frac{\mathrm{d}\mathbb{P}^{\mathrm{R}}\ \thinspace}{\mathrm{d}\mathbb{P}^{\mathrm{R}\nu}}(\omega)$} &
\multirow{2}{*}{$\bar{S}_{\nu}(r(\omega))=-S_{\nu}(\omega)$}\tabularnewline
 &  &  & \tabularnewline
\hline 
\multirow{2}{*}{iDFT} &
\multirow{2}{*}{$\mathbb{P}\left\{ S_{\nu}(\omega)\in\mathrm{d}s\right\} =e^{s}\mathbb{P}^{\nu}\left\{ \tilde{S}_{\nu}(\omega)\in-\mathrm{d}s\right\} $} &
\multirow{2}{*}{$\tilde{S}_{\nu}(\omega)=\ln\frac{\mathrm{d}\mathbb{P}^{\nu}\thinspace\ }{\mathrm{d}\mathbb{P}^{\nu\nu}}(\omega)$} &
\multirow{2}{*}{$\nu$ involutive on $\mathcal{P}$}\tabularnewline
 &  &  & \tabularnewline
\hline 
\multirow{2}{*}{TFT} &
\multirow{2}{*}{$\mathbb{P}\left\{ S_{\nu}(\omega)\in\mathrm{d}s\right\} =e^{s}\mathbb{P}\left\{ S_{\nu}(\omega)\in-\mathrm{d}s\right\} $} &
\multirow{2}{*}{$S_{\nu}(\omega)=\ln\frac{\mathrm{d}\mathbb{P}\ \thinspace}{\mathrm{d}\mathbb{P}^{\nu}}(\omega)$} &
$\nu$ realized by an\tabularnewline
 &  &  & involutive map on $\Omega$\tabularnewline
\hline 
\end{tabular}
\par\end{centering}
\caption{The five fluctuation relations (FRs) of the EP, $S_{\nu}$, with the
CPM operator $\nu$ discussed in this paper. IFT is the integral fluctuation
theorem \cite{jarzynski_nonequilibrium_1997}; GFR is the general
FR in Equation (\ref{eq:GFT}); TFT is the transient fluctuation theorem
in \citep{kurchan_fluctuation_1998,evans_fluctuation_2002,ge_transient_2007,seifert_stochastic_2012},
iDFT is the detailed fluctuation theorem in \citep{esposito_three_2010},
and rDFT is the generalization of Crooks' fluctuation theorem \citep{seifert_stochastic_2012,crooks_entropy_1999,chernyak_path-integral_2006,garcia-garcia_joint_2012}.
We have shown in the text that TFT implies iDFT. This summarizes the
hierarchical structure of the validity of the five FRs. \label{tab:The-five-mainly FR}}
\end{table*}
 Properties of the four EPs we have discussed primarily in this paper
are also summarized in Table \ref{tab:Summary-of-the},
\begin{table*}
\begin{centering}
\begin{tabular}{|l|c|c|c|c|c|c|c|}
\hline 
EP &
$=0$ when &
$\mathbb{E}[\cdot]\ge0$, IFT, GFR &
TFT &
iDFT &
rDFT &
additive in time &
$e^{-\left[\cdot\right]}$ a martingale\tabularnewline
\hline 
\hline 
\multirow{2}{*}{$S_{\mathrm{T}}=\ln\frac{\mathrm{d}\mathbb{P}\ \thinspace}{\mathrm{d}\mathbb{P}^{\mathrm{T}}}$} &
\multirow{2}{*}{TH+SS+DB} &
\multirow{2}{*}{Yes} &
\multirow{2}{*}{Yes} &
\multirow{2}{*}{Yes} &
\multirow{2}{*}{TH+SS} &
\multirow{2}{*}{TH+SS} &
\multirow{2}{*}{TH+SS}\tabularnewline
 &  &  &  &  &  &  & \tabularnewline
\hline 
\multirow{2}{*}{$S_{\mathrm{tot}}=\ln\frac{\mathrm{d}\mathbb{P}\ \ \ }{\mathrm{d}\mathbb{P}^{\mathrm{RT}}}$} &
\multirow{2}{*}{TH+SS+DB} &
\multirow{2}{*}{Yes} &
\multirow{2}{*}{TH+SS} &
\multirow{2}{*}{TH+SS} &
\multirow{2}{*}{if $P_{0}=P_{t}^{\mathrm{R}}$} &
\multirow{2}{*}{Yes} &
\multirow{2}{*}{TH+SS}\tabularnewline
 &  &  &  &  &  &  & \tabularnewline
\hline 
\multirow{2}{*}{$Q_{\mathrm{hk}}=\ln\frac{\mathrm{d}\mathbb{P}\ }{\mathrm{d}\mathbb{P}^{\dagger}}$} &
\multirow{2}{*}{TH+DB} &
\multirow{2}{*}{Yes} &
\multirow{2}{*}{TH+SS} &
\multirow{2}{*}{Yes} &
\multirow{2}{*}{Yes} &
\multirow{2}{*}{Yes} &
\multirow{2}{*}{Yes}\tabularnewline
 &  &  &  &  &  &  & \tabularnewline
\hline 
\multirow{2}{*}{$S_{\mathrm{na}}=\ln\frac{\mathrm{d}\mathbb{P}\ \ \ \ }{\mathrm{d}\mathbb{P}^{\mathrm{R}\dagger\mathrm{T}}}$} &
\multirow{2}{*}{TH+SS} &
\multirow{2}{*}{Yes} &
\multirow{2}{*}{TH+SS} &
\multirow{2}{*}{TH+SS} &
\multirow{2}{*}{if $P_{0}=P_{t}^{\mathrm{R}}$} &
\multirow{2}{*}{Yes} &
\multirow{2}{*}{TH+SS}\tabularnewline
 &  &  &  &  &  &  & \tabularnewline
\hline 
\end{tabular}
\par\end{centering}
\caption{Properties of the four entropy productions (EPs) for Markov Processes.
$S_{\mathrm{T}}$ is the dissipation function \citep{evans_fluctuation_2002,seifert_stochastic_2012};
$S_{\mathrm{tot}}$ is the total\emph{ }EP \citep{harris_fluctuation_2007,esposito_three_2010,seifert_stochastic_2012,ge_physical_2010};
$Q_{\mathrm{hk}}$ is the housekeeping\emph{ }heat \citep{seifert_stochastic_2012,ge_physical_2010}
or called the adiabatic EP in \citep{esposito_three_2010}; and $S_{\mathrm{na}}$
is the non-adiabatic EP \citep{esposito_three_2010}. $\mathbb{E}[\cdot]$
denotes the expectation and the non-negative expectation of entropy
production is the classical second law of thermodynamics. $P_{0}$
is the initial distribution of the original process. $P_{t}^{\mathrm{R}}$
is the terminal distribution of the protocol reversed process, \emph{i.e.
}the distribution one gets by starting with initial distribution $P_{t}$
and then marching with reversed order of transition proability matrices
for $t$ steps. TH stands for time homogeneous; SS stands for steady
state; and DB means the steady state has detailed balance. Note that
when TH and SS, \emph{i.e.} when the system is in nonequilibrium steady
state, we have $S_{\mathrm{T}}\equiv S_{\mathrm{tot}}\equiv Q_{\mathrm{hk}}$
and $S_{\mathrm{na}}=0$. \label{tab:Summary-of-the}}
\end{table*}
. Finally, in Section \ref{Discuss}, we discuss possible future extensions
of our work.

\section{General Theory\label{sec:General-Theories}}

To describe stochastic processes with a measure-theoretic probability
theory, we start by specifying a tuple $\left(\Omega,\mathcal{F}\right)$
called \emph{measurable} \emph{space} where the sample space $\Omega$
collects all possible trajectories $\omega$ and the $\sigma$-algebra
$\mathcal{F}$ collects all events of interest. Physical quantities,
as observables, are then random variables defined on $\left(\Omega,\mathcal{F}\right)$
\citep{qian_mesoscopic_2001}. The statistical properties of a stochastic
process are further specified by a \emph{probability} \emph{space}
$\left(\Omega,\mathcal{F},\mathbb{P}\right)$ with a probability measure
$\mathbb{P}$ that assigns probabilities to events of interest in
$\mathcal{F}$. See Appendix \hyperref{[}{A}{{]}}{A} for more
thorough introduction.

The collection of all possible probability measures on a given measurable
space $\left(\Omega,\mathcal{F}\right)$ forms an affine space of
probability measures $\mathcal{P}$ \citep{hong_representations_2019}.
Each probability measure $\mathbb{P}\in\mathcal{P}$ corresponds to
a stochastic process with specific statistical properties \footnote{Strictly speaking, a stochastic process can be defined without a probability
measure \citep{nutz_pathwise_2012}. However, in this paper, we are
interested in the statistical properties of observables in the process.
We thus say processes with different statistical properties are different
processes.}. In this paper, we would assume $\mathcal{P}$ collects probability
measures that are absolutely continuous to each other (also called
\emph{equivalent} in probability theory), \emph{i.e. }if an event
has zero probability for a stochastic process $\mathbb{P}\in\mathcal{P}$,
the event has zero probability under all $\mathbb{P}$ in $\mathcal{P}$.

\subsection{Change of Probability Measure}

With the statistical properties of stochastic processes specified
by probability measures $\mathbb{P}\in\mathcal{P}$, the difference
between the statistical properties of two processes is characterized
by a change of probability measure (CPM) $\mathbb{P}\rightarrow\mathbb{P}^{\nu}\in\mathcal{P}$,
which induces changes in the statistical properties of observables.
This change in statistical properties can be mathematically represented
by a random variable called the \emph{Radon-Nikodym} \emph{Derivative}
(RND), denoted as $\frac{\mathrm{d}\mathbb{P}^{\nu}}{\mathrm{d}\mathbb{P}\ }(\omega)$
\citep{hong_representations_2019,qian_ternary_2019}. Intuitively,
RND serves as a re-weighting factor in taking expectation. For an
arbitrary random variable $Y(\omega)$ defined on $\left(\Omega,\mathcal{F}\right)$,
it's expectation under $\mathbb{P}^{\nu},$ denoted as $\mathbb{E}^{\nu}[Y(\omega)]$,
can be expressed by the reweighting factor and the previous expectation
$\mathbb{E}[\cdot]$, 
\begin{equation}
\mathbb{E}^{\nu}[Y(\omega)]=\mathbb{E}[Y(\omega)\frac{\mathrm{d}\mathbb{P}^{\nu}}{\mathrm{d}\mathbb{P}\ \thinspace}(\omega)].\label{eq: how RND works}
\end{equation}

Note that to get the probability density function of a random variable
$X(\omega)$, we can let $Y(\omega)$ to be an indicator function
of $X(\omega)$ taking values in between $x$ and $x+\mathrm{d}x$
(denoted as $X(\omega)\in\mathrm{d}x$ for simplicity). That is,
\begin{subequations}
\begin{align}
\mathbb{P}^{\nu}\{X(\omega)\in\mathrm{d}x\} & =\mathbb{E}^{\nu}[\mathbb{I}_{\left\{ \omega:X(\omega)\in\mathrm{d}x\right\} }]\label{eq:-42}\\
 & =\mathbb{E}[\mathbb{I}_{\left\{ \omega:X(\omega)\in\mathrm{d}x\right\} }\frac{\mathrm{d}\mathbb{P}^{\nu}}{\mathrm{d}\mathbb{P}\ \thinspace}(\omega)]\label{eq: pdf of as a expectation}
\end{align}
with the indicator function $\mathbb{I}_{A}(\omega)$ returning 1
if $\omega$ is in the event $A$ and returning 0 otherwise. Throughout
the paper, we use capitalized letters for random variables and their
corresponding lower letters for their values for a specific realization.
\end{subequations}

CPM and RND are key concepts in the theory of fluctuating entropy
and EP \citep{qian_mesoscopic_2001,seifert_entropy_2005,qian_ternary_2019}.
For systems with a discrete sample space $\Omega$ (trajectory space
when considering processes), RND reduces to the ratio of two probability
mass functions, and for those with continuous sample spaces, it reduces
to the ratio of two probability density functions. In either cases,
the RND is obtained from the ratio, which is defined on the codomain
of a random variable, with random variables plugged back in \citep{qian_ternary_2019}.

In stochastic thermodynamics, a physical operation such as a time
reversal in the dynamics is an operation that changes the probability
measure $\mathbb{P}\in\mathcal{P}$ to a new probability measure $\mathbb{P}^{\nu}\in\mathcal{P}$
based on $\mathbb{P}$ alone. This means we are interested in a transformation
$\nu$ on the space of all probability measures $\mathcal{P}\to\mathcal{P}$,
and for each given measure $\mathbb{P}\in\mathcal{P}$, a physical
operation defines a RND that is dependent upon the current process
$\left(\Omega,\mathcal{F},\mathbb{P}\right)$. Thus, we consider operators
$\nu$ that operates on $\mathbb{P}$, giving every measure $\mathbb{P}\in\mathcal{P}$
an image $\mathbb{P}^{\nu}\in\mathcal{P}$ and a corresponding RND.
The new probability measure $\mathbb{P}^{\nu}$ is obtained by 
\begin{equation}
\mathbb{P}^{\nu}\left\{ A\right\} \coloneqq\mathbb{E}\left[\frac{\mathrm{d}\mathbb{P}^{\nu}}{\mathrm{d}\mathbb{P}\ \thinspace}\mathbb{I}_{A}\right],\label{eq: change of measure in explicit form}
\end{equation}
$\forall A\in\mathcal{F}$ and $\forall\mathbb{P}\in\mathcal{P}$.

As we shall show, only very special operator $\nu$ defined on $\mathcal{P}$
can be represented as a result of a map $\mu$ from $\Omega$ to $\Omega$.
In those special cases, the map $\mu$ maps an event of interest $A\in\mathcal{F}$\emph{
}to another, $\mu(A)\in\mathcal{F}$, which is obtained by replacing
all the $\omega$s in $A$ by the $\mu(\omega)$s, \emph{e.g. }if\emph{
$A=\omega_{1}\cup\omega_{2}$}, then $\mu(A)=$\emph{$\mu(\omega_{1})\cup\mu(\omega_{2}).$
}The new measure is then given by 
\begin{equation}
\mathbb{P}^{\nu}\left\{ A\right\} =\mathbb{P}\left\{ \mu(A)\right\} .\label{eq: map induced change of measure}
\end{equation}

\subsection{Fluctuating Entropy Production}

In stochastic thermodynamics, \emph{fluctuating }EP\emph{ }of a CPM
operator $\nu$ is defined as the negative natural logarithm of the
RND, 
\begin{equation}
S_{\nu}(\omega)\coloneqq\ln\frac{\mathrm{d}\mathbb{P}\thinspace\ }{\mathrm{d}\mathbb{P}^{\nu}}(\omega),\label{eq: epr as a -ln RND}
\end{equation}
which is also a random variable \citep{qian_mesoscopic_2001,seifert_entropy_2005,ge_reversibility_2006,ge_generalized_2007,qian_ternary_2019}.
The advantages of working with $S_{\nu}$ instead of the RND can be
seen by its additivity, statistical properties and applications in
information theory \citep{shannon_mathematical_1948,khinchin_mathematical_1957,cover_elements_2006}.
We note that $S_{\nu}$ is always finite given our assumption that
$\mathcal{P}$ collects $\mathbb{P}$s that are absolute continuous
to each other, \emph{i.e.} $0<\frac{\mathrm{d}\mathbb{P}\thinspace\ }{\mathrm{d}\mathbb{P}^{\nu}}(\omega)<\infty$.

The prominent role of the reference measure $\mathbb{P}^{\nu}$ in
the very definition of EP has a clear physical meaning. As the concept
of energy, both entropy and EP are relative to a reference state,
for which the choice is often question dependent. It is well understood
that various different ``free energies'', as thermodynamic potentials,
are determined by the physical settings of an equilibrium ensemble.
In theories of dynamical systems, ergodic stationary measure, with
a translational symmetry in time, has been widely used as a ``natural''
reference in physics and mathematics \citep{young_what_2002}. In
stochastic dynamics, stationarity does not imply local time-reversal
symmetry, which is often coupled to certain parity symmetry. In the
work below, this is best illustrated as an involutive map on $\Omega$
and/or an involutive operation on $\mathcal{P}$.

The EP, $S_{\nu}(\omega)$, can be understood as the fluctuating relative
entropy of trajectories $\omega$ with respect to the reference probability
measure $\mathbb{P}^{\nu}.$ It reflects the difference between two
probability measures. If the stochastic process is symmetric under
the operator $\nu$, \emph{i.e.} $\mathbb{P}^{\nu}=\mathbb{P}$, then
$S_{\nu}(\omega)=0,\text{\ensuremath{\forall\omega\in\Omega}}$. With
different $\nu$, we can have various different EPs that are physically
important, \emph{e.g. }the nonadiabatic EP that is related to work
and heat. It is therefore desirable to find the general statistical
properties of a given EP given it's definition in Equation (\ref{eq: epr as a -ln RND}).

\subsection{Fluctuation Relations}

Directly from the definition of EP in Equation (\ref{eq: epr as a -ln RND}),
the following three key statistical properties of $S_{\nu}$ can be
derived rather straightforwardly.
\begin{flushleft}
(a) \emph{Non-negative expectation:} By Jensen's inequality, the expectation
of $S_{\nu}$ w.r.t $\mathbb{P}$ is non-negative, $\mathbb{E}[S_{\nu}]\ge0$,
and equality only holds when $\mathbb{P}=\mathbb{P}^{\nu}$ due to
the strict convexity of negative logarithm. This result for EPs in
physical processes extends the classical second law of thermodynamics
\citep{ge_extended_2009}.
\par\end{flushleft}

\begin{flushleft}
(b) \emph{Integral Fluctuation Theorem} (IFT) or called Jarzynski's
equality \citep{jarzynski_nonequilibrium_1997}: 
\begin{equation}
\mathbb{E}\left[e^{-S_{\nu}(\omega)}\right]=\mathbb{E}\left[\frac{\ \mathrm{d}\mathbb{P}^{\nu}}{\mathrm{d}\mathbb{P}}(\omega)\right]=\mathbb{E}^{\nu}[1]=1.\label{eq: IFT}
\end{equation}
\par\end{flushleft}

\begin{flushleft}
(c) \emph{General fluctuation relation} (GFR): 
\begin{equation}
\mathbb{P}\left\{ S_{\nu}(\omega)\in\mathrm{d}s\right\} =e^{s}\mathbb{P}^{\nu}\left\{ S_{\nu}(\omega)\in\mathrm{d}s\right\} \label{eq: general detail FT}
\end{equation}
where $\mathrm{d}s$ is a shorthand for the infinitesimal interval
$\left(s,s+\mathrm{d}s\right)$. This GFR states for the EP, $S_{\nu}=\ln\frac{\mathrm{d}\mathbb{P}\ }{\mathrm{d}\mathbb{P}^{\nu}}$,
that quantifies the difference between $\mathbb{P}$ and $\mathbb{P}^{\nu}$,
its probability densities under $\mathbb{P}$ and $\mathbb{P}^{\nu}$
are up to a exponential factor.
\par\end{flushleft}

GFR can be derived by considering the probability density of $S_{\nu}$
under the new measure $\mathbb{P}^{\nu}$, characterizing the statistical
properties of $S_{\nu}$ in the $\nu$ process:
\begin{subequations}
\begin{align}
\mathbb{P}^{\nu}\left\{ S_{\nu}(\omega)\in\mathrm{d}s\right\}  & =\mathbb{E}\left[\frac{\mathrm{d}\mathbb{P}^{\nu}}{\mathrm{d}\mathbb{P}\thinspace\ }\mathbb{I}_{\left\{ \omega:S_{\nu}(\omega)\in\mathrm{d}s\right\} }\right]\label{eq:-43}\\
 & =\mathbb{E}\left[e^{-S_{\nu}(\omega)}\mathbb{I}_{\left\{ \omega:S_{\nu}(\omega)\in\mathrm{d}s\right\} }\right]\label{eq:-55}\\
 & =e^{-s}\mathbb{P}\left\{ S_{\nu}(\omega)\in\mathrm{d}s\right\} .\label{eq:GFT}
\end{align}
We will see below that most fluctuation relations discussed in the
literature come directly from this GFR. We remark that the three properties
above holds for \emph{any} fluctuating relative entropy defined as
the negative logarithm of a RND.
\end{subequations}

\subsubsection{Detailed Fluctuation Theorems}

The detailed fluctuation theorems that were considered in the literature
\citep{crooks_entropy_1999,chernyak_path-integral_2006,esposito_three_2010,seifert_stochastic_2012}
have the form of 
\begin{equation}
\mathbb{P}\left\{ S_{\nu}(\omega)\in\mathrm{d}s\right\} =e^{s}\mathbb{P}^{\nu}\left\{ \hat{S}_{\nu}(\omega)\in-\mathrm{d}s\right\} \label{eq: general form of DFT}
\end{equation}
where $\hat{S}_{\nu}$ can be various different random variables under
different considerations. The DFT comes directly from GFR in Equation
(\ref{eq:GFT}) if there is an odd parity between the original random
variable $S_{\nu}$ and the new random variable $\hat{S}_{\nu}$ under
consideration, 
\begin{equation}
\hat{S}_{\nu}(\omega)=-S_{\nu}(\omega).\label{eq: odd parity}
\end{equation}
Two choices of $\hat{S}_{\nu}$ were discussed in the past, which
we will briefly summarize below.

Recall that $S_{\nu}$ serves as a random variable that quantifies
the effect of the CPM operator $\nu$ acting on $\left(\Omega,\mathcal{F},\mathbb{P}\right)$.
When considering a different process $\left(\Omega,\mathcal{F},\mathbb{P}^{\eta}\right)$,
the EP that does the same for the $\eta$ process as $S_{\nu}$ does
for the original process should be given by replacing $\mathbb{P}$
with $\mathbb{P}^{\eta}$, 
\begin{equation}
S_{\eta\nu}^{\eta}(\omega)\coloneqq\ln\frac{\mathrm{d}\mathbb{P}^{\eta}\ \thinspace}{\mathrm{d}\mathbb{P}^{\eta\nu}}(\omega)\label{eq: definition change for S after change of measure}
\end{equation}
where $\mathbb{P}^{\eta\nu}=\left[\mathbb{P}^{\eta}\right]^{\nu}$
is operating $\eta$ on $\mathbb{P}$ first and then applying $\nu$
on $\mathbb{P}^{\eta}$. The two $\hat{S}_{\nu}$s considered in the
past \citep{esposito_three_2010,seifert_stochastic_2012} correspond
to two different $\eta$s.

A mathematically natural consideration for $\eta$ is to take $\eta$
as $\nu$, which will lead to Esposito and Van den Broeck's detailed
fluctuation theorem in \citep{esposito_three_2010}. In this setting,
the odd parity requirement in Equation (\ref{eq: odd parity}) becomes
an involutive requirement of the operator $\nu$, 
\begin{equation}
S_{\nu\nu}^{\nu}=\ln\frac{\mathrm{d}\mathbb{P}^{\nu}\ \thinspace}{\mathrm{d}\mathbb{P}^{\nu\nu}}=-S_{\nu}=\ln\frac{\mathrm{d}\mathbb{P}^{\nu}}{\mathrm{d}\mathbb{P}\ }\Leftrightarrow\mathbb{P}^{\nu\nu}=\mathbb{P}.\label{eq: involutive requirement}
\end{equation}
Denoting $S_{\nu\nu}^{\nu}$ as $\tilde{S}_{\nu}$, the detailed fluctuation
theorem from the involutive property (iDFT) \citep{esposito_three_2010}
then reads 
\begin{equation}
\mathbb{P}\left\{ S_{\nu}(\omega)\in\mathrm{d}s\right\} =e^{s}\mathbb{P}^{\nu}\left\{ \tilde{S}_{\nu}(\omega)\in-\mathrm{d}s\right\} .\label{eq: iDFT}
\end{equation}

In a physical process, the driving protocol of the system is determined
by macroscopic thermodynamics parameters, and thermodynamics quantities
such as heat and work are dependent upon the driving protocol. Therefore,
in most of the physics literature, the $\hat{S}_{\nu}$ considered
was given by first taking $\eta$ as the macroscopic, protocol reversal
$\mathrm{R}$, as we will defined explicitly in Equation (\ref{eq: joint for  protocol reversal R}),
and then evaluating $\hat{S}_{\nu}$ at the order reversed trajectory
$r(\omega)$ where $r$ is a map from $\Omega$ to $\Omega$ that
reverses the trajectory $\omega\in\Omega.$ By denoting $\bar{S}_{\nu}(\omega)\coloneqq S_{\mathrm{R}\nu}^{\mathrm{R}}(\omega)$
and using $\bar{S}_{\nu}(r(\omega))$ for $\hat{S}(\omega)$, we get
the generalized Crooks' fluctuation theorem \citep{crooks_entropy_1999,crooks_path-ensemble_2000,chernyak_path-integral_2006,seifert_stochastic_2012,garcia-garcia_joint_2012},
\begin{equation}
\mathbb{P}\left\{ S_{\nu}(\omega)\in\mathrm{d}s\right\} =e^{s}\mathbb{P}^{\mathrm{\nu}}\left\{ \bar{S}_{\nu}(r(\omega))\in-\mathrm{d}s\right\} .\label{eq: rDFT}
\end{equation}
To fix the terminology, we would refer this detailed fluctuation theorem
as the rDFT. The odd parity requirement $\bar{S}_{\nu}(r(\omega))=-S_{\nu}(\omega)$
turns out to be an involutive requirement on the operator $\mathrm{R}$
for the CPM operators we are interested in, as shown in Equation (\ref{eq: rDFT for Stot})
and (\ref{eq: P0 =00003D Pt^R requirement for Sna}).

To check the validity of iDFT and rDFT, we should check the necessary
and sufficient condition: for $S_{\nu}$ and $-\hat{S}_{\nu}$ to
have the same probability density of under $\mathbb{P}^{\nu}$. The
odd parity condition in Equation (\ref{eq: odd parity}) is a stricter
condition on $S_{\nu}$ and $\hat{S}_{\nu}$ \footnote{Note that saying two random variables to be the same $X(\omega)=Y(\omega)$
usually means that they return the same value for every $\omega\in\Omega$.
This is a stronger statement than saying two random variables $X(\omega)$
and $Y(\omega)$ to have the same probability distribution.} and only a sufficient condition.. To prove a DFT to be valid, we
can show this sufficient condition to be true, but to prove a DFT
to be invalid, we would need to show a necessary condition of it to
be false. For simplicity, we will only check the the sufficient condition
in Equation (\ref{eq: odd parity}) for a DFR in this paper. If an
EP does not admit this odd-parity condition, we will leave its DFT
inconclusive and leave it for future consideration.

\subsubsection{Transient Fluctuation Theorem}

If an operator $\nu$ is involutive, \emph{i.e. $\mathbb{P}^{\nu\nu}=\mathbb{P}$},
then we already know that $S_{\nu}$ admits iDFT. Now, if the operator
$\nu$ is further realized by an involutive map $\mu:\Omega\rightarrow\Omega$
on the trajectory space as we have shown in Equation (\ref{eq: map induced change of measure}),
\emph{i.e. }$\mu(\mu(\omega))=\omega$, we would have $S_{\nu}(\mu(\omega))=\ln\frac{\mathrm{d}\mathbb{P}^{\nu}\ }{\mathrm{d}\mathbb{P}^{\nu\nu}}(\omega)=\tilde{S}_{\nu}(\omega)=\ln\frac{\mathrm{d}\mathbb{P}^{\nu}}{\mathrm{d}\mathbb{P}\ }(\omega)=-S_{\nu}(\omega)$.
Then, by Equation (\ref{eq: map induced change of measure}), we have
$\mathbb{P}^{\nu}\left\{ S_{\nu}(\omega)\in\mathrm{d}s\right\} =\mathbb{P}\left\{ S_{\nu}(\mu(\omega))\in\mathrm{d}s\right\} $.
With our GFR, we obtain the so-called \emph{transient} \emph{fluctuation}
\emph{theorem }(TFT) \citep{evans_equilibrium_1994,searles_fluctuation_1999,evans_fluctuation_2002,jiang_mathematical_2004,ge_transient_2007,seifert_stochastic_2012},
\begin{equation}
\mathbb{P}\left\{ S_{\nu}(\omega)\in\mathrm{d}s\right\} =e^{s}\mathbb{P}\left\{ S_{\nu}(\omega)\in-\mathrm{d}s\right\} .\label{eq: transient fluctuation theorem}
\end{equation}
TFT is particularly important since it provides explicitly the asymmetry
between having a positive and negative EP in the same process. The
probability (density) of finding a positive EP is exponentially higher
than the probability of finding a negative one.

The validity of TFT is easy to check since involutive property $\mathbb{P}^{\nu\nu}=\mathbb{P}$
is an necessary condition for TFT. If $\mathbb{P}^{\nu\nu}\neq\mathbb{P}$
then $\mathbb{E}[\ln\frac{\mathrm{d}\mathbb{P}\ \ }{\mathrm{d}\mathbb{P}^{\nu\nu}}]>0\Rightarrow\mathbb{E}[S_{\nu}]>\mathbb{E}[-S_{\nu\nu}^{\nu}]$
by Jensen's inequality. We then know $S_{\nu}$ and $-S_{\nu\nu}^{\nu}$
have different probability densities w.r.t. $\mathbb{P}$ $\Rightarrow$
TFT is false. Here, we see that TFT is a sufficient condition for
$S_{\nu}$ to have iDFT but not necessary. This is because not all
involutive operator $\nu:\mathcal{P}\rightarrow\mathcal{P}$ can be
realized by an involutive map $\mu:\Omega\rightarrow\Omega$ \footnote{\textit{\emph{Here is an explicit example for an involutive operator
$\nu:\mathcal{P}\rightarrow\mathcal{P}$ not realized by an involutive
map $\mu:\Omega\rightarrow\Omega$. Consider two binary random variables
$X_{1}$ and $X_{2}$ that can take values $0$ or $1$. Suppose the
joint probabilities for four possible realizations are $P_{1,2}(0,0)=p,P_{1,2}(0,1)=q,P_{1,2}(1,0)=r,P_{1,2}(1,1)=s$
where $p+q+r+s=1$. The joint probabilities in the new measure $\mathbb{P}^{\nu}$
is given by $P_{1,2}^{\nu}(0,0)=\sqrt{\left(1-r-s\right)-p^{2}},P_{1,2}^{\nu}(0,1)=\left(1-r-s\right)-\sqrt{\left(1-r-s\right)-p^{2}},P_{1,2}^{\nu}(1,0)=r,P_{1,2}^{\nu}(1,1)=s$.
One can show that $\mathbb{P}^{\nu\nu}=\mathbb{P}$. And clearly the
involutive operator $\nu$ can not be realized by an involutive map
$\mu:\Omega\rightarrow\Omega$.}}}\textit{\emph{.}}

\subsection{Summary}

Treating a physical operation on stochastic processes as a CPM operator
$\nu$ on probability space $\mathcal{P}$, we have characterized
the change in the statistical properties of a physical operation via
the negative natural logarithm of the RND, which one defines it as
the EP $S_{\nu}$. In fact, a hierarchy of the validity for FRs in
general stochastic processes is revealed from our work as summarized
in Table \ref{tab:The-five-mainly FR}:
\begin{flushleft}
(a) Non-negative expectation, IFT \citep{jarzynski_nonequilibrium_1997,seifert_stochastic_2012},
and GFR are generally true from the definition of $S_{\nu}$.
\par\end{flushleft}

\begin{flushleft}
(b) With $\hat{S}_{\nu}(\omega)=-S_{\nu}(\omega)$, we have DFTs.
In the literature, $\hat{S}_{\nu}(\omega)$ was chosen to be $\tilde{S}_{\nu}(\omega)\coloneqq S_{\nu\nu}^{\nu}(\omega)$
to get iDFT \citep{esposito_three_2010} if the CPM operator $\nu$
is involutive, or $\bar{S}_{\nu}(r(\omega))\coloneqq S_{\mathrm{R}\nu}^{\mathrm{R}}(r(\omega))$
to get rDFT \citep{crooks_entropy_1999,crooks_path-ensemble_2000,seifert_stochastic_2012}
if the protocol reversal operator $\mathrm{R}$ is involutive (for
the $\nu$ we considered).
\par\end{flushleft}

\begin{flushleft}
(c) Further with the CPM operator $\nu$ as an involutive map on the
trajectory space, from $\Omega\rightarrow\Omega$, we have TFT \citep{evans_fluctuation_2002,ge_transient_2007,seifert_stochastic_2012}.
\par\end{flushleft}

The results above are true no matter the stochastic process has discrete
or continuous state space $\mathcal{X}$, is with discrete or continuous
time, is time homogeneous or not, or has any specific initial distribution
such as the invariant distribution. The Markovian assumption is not
even imposed except the definition of the protocol reversal $\mathrm{R}$.
Our derivation only relies on assuming all $\mathbb{P}\in\mathcal{P}$
to be\emph{ }absolute\emph{ }continuous\emph{ }to\emph{ }each\emph{
}other, the notation of CPM operator, the definition of EP, and conditions
for more restricted FRs such as DFTs and TFT.

With these general results in hand, we shall consider EPs in physics
and chemistry, the reversal operators they correspond to, and their
fluctuation relations as examples in following sections. We will start
by introducing different reversals in Section \ref{sec:Different-Types-of-reversals}
and then various EPs with their fluctuation relations in Section \ref{sec:Entropy-Productions-in Physics and Chemistry}.
As we will see, our CPM operator notion clarifies the difference between
different time reversals and between EPs, especially between the dissipation
function and the total entropy production, which are easily confused
quantities.

\section{Different Types of Reversal\label{sec:Different-Types-of-reversals}}

EPs in nonequilibrium physics and chemistry are introduced by comparing
the original process to its ``time reversal'' \citep{seifert_stochastic_2012}.
The definition of a time reversal, however, is inevitably based on
our understanding of the physics of time. See \citep{qian_zeroth_2014}
for a discussion of ``overdamped'' vs. ``underdamped'' thermodynamics.

Here, for general Markov processes, we would consider three different
reversals. We use a discrete-time Markov chain with $t$ time steps
and discrete state space $\mathcal{X}$ as a paradigm. Markov processes
in continuous time and continuous space will be discussed in Section
\ref{sec:Entropy-Productions-in Diffusion}. 

We use the colon notation $X_{0:t}$ to represent a sequence of random
variables $\left(X_{0},X_{1},\cdots,X_{t}\right)$ and a specific
trajectory $x_{0:t}=\left(x_{0},x_{1},x_{2},\cdots,x_{t}\right)$.
In our consideration, $\omega$ is a specific trajectory $x_{0:t}$
and our trajectory space $\Omega$ is given by the outer product of
$\left(t+1\right)$ state spaces, $\mathcal{X}\otimes\mathcal{X}\otimes\cdots\otimes\mathcal{X}$
or simply $\mathcal{X}^{t+1}$. The full probabilistic description
of the state variables $X_{0:t}$ is given by their joint probability
denoted as 
\begin{equation}
P_{0:t}(x_{0:t})\coloneqq\mathbb{P}\left\{ X_{0}=x_{0},X_{1}=x_{1},\cdots,X_{t}=x_{t}\right\} .\label{eq:}
\end{equation}
The marginal probabilities $P_{n}(x_{n})\coloneqq\mathbb{P}\left\{ X_{n}=x_{n}\right\} $
and the conditional probabilities $P_{m|n}(x_{m}|x_{n})\coloneqq\mathbb{P}\left\{ X_{m}=x_{m}|X_{n}=x_{n}\right\} $
can be computed from the joint probability. We would denote the transition
matrix at the $n$th time step as 
\begin{equation}
M_{n}(x_{n}|x_{n-1})\coloneqq P_{n|n-1}(x_{n}|x_{n-1}).\label{eq:-1}
\end{equation}
With this notion, the joint probability for a time-inhomogeneous Markov
process, 
\begin{equation}
P_{0:t}(x_{0:t})=P_{0}(x_{0})\prod_{n=1}^{t}M_{n}(x_{n}|x_{n-1}),\label{eq:-3}
\end{equation}
is determined by the \emph{driving} \emph{protocol}, which constitutes
the initial distribution $P_{0}$ and all the transition matrices
$M_{n}$ for $n=1,2,\cdots,t$. For each transition matrix, we also
assumed the existence of a unique invariant distribution $\pi_{n}$
satisfying $\sum_{i\in\mathcal{X}}\pi_{n}(i)M_{n}(j|i)=\pi_{n}(j).$

\subsection{Time Reversal}

The time reversal of a Markov Chain for $n=0,1,\cdots,t$ is conventionally
defined by a change of random variable 
\begin{equation}
X_{n}^{\mathrm{T}}(\omega)=X_{t-n}(\omega)\label{eq:-4}
\end{equation}
where we use superscript $\mathrm{T}$ to represent time reversal.
This definition of $X_{n}^{\mathrm{T}}$ can be treated as the random
variable induced by a map on the trajectory space, $r:\Omega\rightarrow\Omega$,
\begin{equation}
X_{n}^{\mathrm{T}}(\omega)=X_{n}(r(\omega)),\label{eq:-5}
\end{equation}
where the map $r$ reverses the order of a trajectory $\omega$, $r(x_{0:t})=x_{t:0}$.
Given a specific trajectory $\omega=x_{0:t}$, the state variable
$X_{n}(\omega)$ is understood as the observed state of the system
at time $n$. We can then clearly see the equivalence between these
two definitions, 
\begin{equation}
X_{n}^{\mathrm{T}}(x_{0:t})=X_{t-n}(x_{0:t})=x_{t-n}=X_{n}(x_{t:0}).\label{eq: two def for TR RV}
\end{equation}

By the equivalence between change of random variable and change of
probability measure \citep{qian_ternary_2019}, instead of regarding
time reversal as a change of random variable, we can also characterize
the time reversal as a change of probability measure with a CPM operator
$\mathrm{T}$. The CPM operator $\mathrm{T}$ is realized by the map
$r:\Omega\rightarrow\Omega$ on the trajectory space. The joint probability
after time reversal is thus given by
\begin{subequations}
\begin{align}
P_{0:t}^{\mathrm{T}}(x_{0:t}) & =\mathbb{P}^{\mathrm{T}}\{X_{0}(\omega)=x_{0},\cdots,X_{t}(\omega)=x_{t}\}\\
 & =\mathbb{P}\{X_{0}(r(\omega))=x_{0},\cdots X_{t}(r(\omega))=x_{t}\}\\
 & =\mathbb{P}\{X_{t}(\omega)=x_{0},\cdots,X_{0}(\omega)=x_{t}\}\\
 & =P_{t:0}(x_{0:t})\\
 & =\mathbb{P}\{X_{0}(\omega)=x_{t},\cdots X_{t}(\omega)=x_{0}\}\label{eq: joint for TR}\\
 & =P_{0:t}(x_{t:0}).
\end{align}
We see that the joint probability of finding $x_{0:t}$ in the time
reversed process is the same as the joint probability of finding the
order-reversed trajectory $x_{t:0}$ in the original process. The
assumption that the order-reversed trajectory has a nonzero probability
in the original process is the \emph{microscopic reversible} assumption
required in \citep{crooks_entropy_1999}.
\end{subequations}

When using a change of probability measure perspective, the meaning
of the random variables $X_{n}$ is preserved as the $n$th state
of the process. The changes in its statistical properties are due
to the change in probability measure. An interesting analog to these
two equivalence ways of characterizing the change is the Schrödinger's
and Heisenberg's pictures of quantum mechanics \citep{qian_ternary_2019}.
We also note that, from the results above, it can be mathematically
shown that the time reversed Markov chain is still Markovian but will
be time inhomogeneous even if the original process is time homogeneous.

\subsection{Protocol Reversal}

The joint probability of a Markov Chain is determined by the \emph{driving}
\emph{protocol,} $P_{0}$ and $M_{n}$, $n=1,2,\cdots,t$. Thus, we
can consider the process where we used the terminal distribution $P_{t}$
as our new initial distribution and reverse the temporal order of
the transition matrices. We shall call this reversal the \emph{protocol
reversal} of the process and denote the corresponding CPM operator
as $\mathrm{R}$. The new joint distribution is then given by 
\begin{equation}
P_{0:t}^{\mathrm{R}}(x_{0:t})=P_{t}(x_{0})\prod_{n=1}^{t}M_{t+1-n}(x_{n}|x_{n-1}).\label{eq: joint for  protocol reversal R}
\end{equation}
Compare to the time reversal $\mathrm{T}$ which is a ``time reversal''
at the microscopic/trajectory level, protocol reversal $\mathrm{R}$
is rather a ``time reversal'' at the macroscopic/thermodynamics
level.
\begin{subequations}
The ``time reversal'' that was considered by most of the previous
studies on fluctuation relations \citep{seifert_entropy_2005,chernyak_path-integral_2006,esposito_three_2010}
is in fact the composition of the two reversals we have just introduced:
$\mathrm{R}$ and $\mathrm{T}$, denoted as\emph{ $\mathbb{P}^{\mathrm{R}\mathrm{T}}$.
}The joint distribution is given by

\emph{
\begin{align}
\left[P_{0:t}^{\mathrm{R}}\right]^{\mathrm{T}}(x_{0:t}) & =P_{0:t}^{\mathrm{R}}(x_{t:0})\\
 & =P_{t}(x_{t})\prod_{n=1}^{t}M_{t+1-n}(x_{t-n}|x_{t-n+1})\label{eq: R and T composition}
\end{align}
}
\end{subequations}

This computation in Equation (\ref{eq: R and T composition}) actually
gives us a convenient result when working on composite CPMs with time
reversal as the last operation, $\mathbb{P}\rightarrow\mathbb{P}^{\nu\mathrm{T}}$.
The joint probability for such composite operators is given by evaluating
at the order-reversed trajectory, 
\begin{equation}
P_{0:t}^{\nu\mathrm{T}}(x_{0:t})=P_{0:t}^{\nu}(x_{t:0}).\label{eq:-56}
\end{equation}
Note that the two operators $\mathrm{R}$ and $\mathrm{T}$ do not
generally commute, $P_{0:t}^{\mathrm{RT}}\neq P_{0:t}^{\mathrm{TR}}$\footnote{From definition, $P_{0:t}^{\mathrm{TR}}(x_{0:t})=\left[P_{0:t}^{\mathrm{T}}\right]^{\mathrm{R}}(x_{0:t})$
is $P_{t}^{\mathrm{T}}(x_{0})\prod_{n=1}^{t}P_{t+1-n|t-n}^{\mathrm{T}}(x_{n}|x_{n-1})$.
From the fact that $\mathrm{T}$ changes the random variable, we get
the RHS equals to $P_{0}(x_{0})\prod_{n=1}^{t}P_{n-1|n}(x_{n}|x_{n-1})$
and becomes $\frac{P_{0}(x_{0})}{P_{t}(x_{t})}\prod_{n=1}^{t}\frac{P_{n-1}(x_{n})}{P_{n}(x_{n-1})}P_{0:t}^{\mathrm{RT}}(x_{0:t})$
by Bayes' rule.}.

\subsection{the Dual Process}

The last reversal we consider in this paper is by introducing the
driving protocol that reverses the probability flux in the invariant
steady state \emph{at} \emph{each} \emph{time} \emph{step}. This new
process is called the \emph{dual} process \citep{seifert_stochastic_2012,crooks_path-ensemble_2000}
of the original process. For a time homogeneous process, the dual
process is equivalent to the time reversal of the process if the process
starts and stays in the invariant steady state. However, for a general
time inhomogeneous process, the correspondence between the dual and
the time reversal can only be drawn within each given time step.

For the $n$th time step where the transition matrix is $M_{n}$ and
the invariant distribution is $\pi_{n}$, the probability flux from
state $i$ to state $j$ is given by the joint probability difference
between $i\rightarrow j$ and $j\rightarrow i$, 
\begin{equation}
J_{n}(i,j)\coloneqq\pi_{n}(i)M_{n}(j|i)-\pi_{n}(j)M_{n}(i|j).\label{eq:-7}
\end{equation}
The probability flux can then be reversed, $J_{n}^{\dagger}(i,j)=-J_{n}(i,j)$,
by replacing $M_{n}$ with its dual matrix 
\begin{equation}
M_{n}^{\dagger}(j|i)=\frac{\pi_{n}(j)}{\pi_{n}(i)}M_{n}(i|j).\label{eq: the dual transition matrix}
\end{equation}
 The definition of a dual process is thus given by replacing all the
$M_{n}$ with $M_{n}^{\dagger},$ 
\begin{equation}
P_{0:t}^{\dagger}(x_{0:t})=P_{0}(x_{0})\prod_{n=1}^{t}M_{n}^{\dagger}\left(x_{n}|x_{n-1}\right).\label{eq: the dual process}
\end{equation}
It can be shown from Equation (\ref{eq: the dual transition matrix})
that $M_{n}$ and $M_{n}^{\dagger}$ have the same invariant distribution
$\pi_{n}$.

Recall the detailed balance condition is given by 
\begin{equation}
\pi_{n}(i)M_{n}(j|i)=\pi_{n}(j)M_{n}(i|j),\forall i,j\in\mathcal{X},\label{eq:-8}
\end{equation}
which is equivalent to $J_{n}^{\dagger}=J_{n}=0$ and $M_{n}^{\dagger}=M_{n}$.
Therefore, comparing the dual process to the original one directly
reveals whether the system possesses detailed balance or not. Detailed
balance systems are invariant under the CPM operator $\dagger$.

\subsection{Involutive Properties of the Reversals}

Considering reversals of a process, it is natural to ask whether we
can recover the original process by applying the reversal twice or
not, \emph{i.e.} in mathematical terms, whether the operator is involutive
or not. As we have shown above, the involutive properties of the CPM
operator for a EP are the keys for the EP to have FRs.

It is rather straightforward to show that both $\mathrm{T}$ and $\dagger$
are involutive. The time reversal $\mathrm{T}$ is involutive since
the map $r:\Omega\rightarrow\Omega$ is involutive. We can verify
this by computing $P_{0:t}^{\mathrm{TT}}(x_{0:t})=P_{0:t}^{\mathrm{T}}(x_{t:0})=P_{0:t}(x_{0:t}).$
The dual reversal $\dagger$ is involutive by computing $P_{0:t}^{\dagger\dagger}(x_{0:t})=P_{0}^{\dagger}(x_{0})\prod_{n=1}^{t}M_{n}^{\dagger\dagger}(x_{n}|x_{n-1}).$
From the joint probability in Equation (\ref{eq: the dual process}),
we get $P_{0}^{\dagger}=P_{0}$ and one can show $M_{n}^{\dagger\dagger}=M_{n}$
by $\pi_{n}^{\dagger}=\pi_{n}$.

Finally, the protocol reversal $\mathrm{R}$ is \emph{not} involutive
in general. To see this, we start by 
\begin{align}
P_{0:t}^{\mathrm{RR}}(x_{0:t})= & P_{t}^{\mathrm{R}}(x_{0})\prod_{n=1}^{t}M_{t+1-n}^{\mathrm{R}}(x_{n}|x_{n-1}).
\end{align}
We thus need to compute $P_{t}^{\mathrm{R}}$ and $M_{t+1-n}^{\mathrm{R}}$
from the joint probability given in Equation (\ref{eq: joint for  protocol reversal R}).
It is straightforward to check that the latter is given simply by
\begin{equation}
M_{t+1-n}^{\mathrm{R}}(j|i)=M_{n}(j|i).\label{eq:-11}
\end{equation}
However, the terminal distribution of the protocol reversed process
is generally not the initial distribution of the original process,
$P_{t}^{\mathrm{R}}\neq P_{0}$ \citep{rao_detailed_2018}. This can
be seen by a time homogeneous Markov Chain where $P_{t}^{\mathrm{R}}$
is given by $P_{t}$ further evolved by $t$ more steps, which gives
us $P_{2t}$ not $P_{0}$. Thus, we have

\begin{align}
P_{0:t}^{\mathrm{RR}}(x_{0:t}) & =\frac{P_{t}^{\mathrm{R}}(x_{0})}{P_{0}(x_{0})}P_{0:t}(x_{0:t}).\label{eq: involutive R}
\end{align}
From this, it is also clear that if $P_{t}^{\mathrm{R}}=P_{0}$, then
the protocol reversal $\mathrm{R}$ becomes involutive \footnote{An example for an involutive $\mathrm{R}$ is to start at $P_{0}=\pi_{1}$
and to fix the driving protocol at $M_{1}$ for enough time steps
so that the protocol reversed process have enough time to relax back
to $\pi_{1}$ by the end of the reversed process.}.

\section{Entropy Productions in Physics and Chemistry\label{sec:Entropy-Productions-in Physics and Chemistry}}

With different reversals and their corresponding CPM operators introduced,
we are now ready to consider different EPs in physics and chemistry
and their fluctuation relations. We already knew that every EP as
a fluctuating relative entropy has a non-negative expectation and
admit both IFT and GFR. Thus, we would mainly discuss the rDFT, iDFT,
and TFT for various different EPs in physics and chemistry.

\subsection{Dissipation Function $S_{\mathrm{T}}$}

The EP that corresponds to the time reversal $\mathrm{T}$ is historically
called the \emph{dissipation} \emph{function }by Evans and Searles
\citep{seifert_stochastic_2012,evans_fluctuation_2002}, a term goes
back to Onsager, 
\begin{equation}
S_{\mathrm{T}}(\omega)\coloneqq\ln\frac{\mathrm{d}\mathbb{P}\ \thinspace}{\mathrm{d}\mathbb{P}^{\mathrm{T}}}(\omega)=\ln\frac{P_{0:t}(X_{0:t}(\omega))}{P_{0:t}(X_{t:0}(\omega))}.\label{eq:dissipation function}
\end{equation}
We note that the dissipation function does \emph{not} satisfy additive
properties when connecting two time intervals, \emph{i.e. }for $0<s<t$,
we have\emph{ 
\begin{equation}
S_{\mathrm{T}}(x_{0:t})\neq S_{\mathrm{T}}(x_{0:s})+S_{\mathrm{T}}(x_{s:t})\label{eq: non-additivitiy in ST}
\end{equation}
}

Since the CPM operator $\mathrm{T}$ is realized by an involutive
map $r:\Omega\rightarrow\Omega$. We thus know $S_{\mathrm{T}}$ admits
both TFT and iDFT. The TFT of $S_{\mathrm{T}}$ has been discussed
in \citep{seifert_stochastic_2012,ge_transient_2007,evans_fluctuation_2002}.
However, $S_{\mathrm{T}}$ does not satisfy the odd-parity, sufficient
condition for rDFT. We note that the dissipation function $S_{\mathrm{T}}$
on the protocol reversed process $\mathrm{R}$ is given by $\bar{S}_{\mathrm{T}}(\omega)=\ln\frac{\mathrm{d}\mathbb{P}^{\mathrm{R}}\ \thinspace}{\mathrm{d}\mathbb{P}^{\mathrm{RT}}}(\omega)$
which gives
\begin{subequations}
\begin{align}
\bar{S}_{\mathrm{T}}(r(\omega)) & =\ln\frac{\mathrm{d}\mathbb{P}^{\mathrm{RT}}\ \thinspace}{\mathrm{d}\mathbb{P}^{\mathrm{RTT}}}(\omega)=\ln\frac{\mathrm{d}\mathbb{P}^{\mathrm{RT}}}{\mathrm{d}\mathbb{P}^{\mathrm{R}}\ \thinspace}(\omega)\\
 & =-\bar{S}_{\mathrm{T}}(\omega)\neq-S_{\mathrm{T}}(\omega).
\end{align}
Unless pathologically $\bar{S}_{\mathrm{T}}(r(\omega))\neq-S_{\mathrm{T}}(\omega)$
but $\mathbb{P}^{\mathrm{\mathrm{T}}}\left\{ \bar{S}_{\mathrm{T}}(r(\omega))\in-\mathrm{d}s\right\} =\mathbb{P}^{\mathrm{T}}\{S_{\mathrm{T}}(\omega)\in-\mathrm{d}s\}$,
the dissipation function $S_{\mathrm{T}}$ would not admit rDFT.
\end{subequations}

\subsection{Total Entropy Production $S_{\mathrm{tot}}$}

The \emph{total} \emph{entropy production }discussed in \citep{harris_fluctuation_2007,ge_physical_2010,seifert_stochastic_2012,esposito_three_2010}
is given by composing protocol reversal $\mathrm{R}$ and then the
time reversal $\mathrm{T}$, \begin{subequations}
\begin{align}
S_{\mathrm{tot}}(x_{0:t}) & =\ln\frac{\mathrm{d}\mathbb{P}\ \ \ }{\mathrm{d}\mathbb{P}^{\mathrm{RT}}}(x_{0:t})=\ln\frac{\mathrm{d}\mathbb{P}\ \thinspace}{\mathrm{d}\mathbb{P}^{\mathrm{C}}}(x_{0:t})\label{eq:-10}\\
 & =\ln\frac{P_{0}(x_{0})M_{1}(x_{1}|x_{0})\cdots M_{t}\left(x_{t}|x_{t-1}\right)}{P_{t}(x_{t})M_{t}(x_{t-1}|x_{t})\cdots M_{1}\left(x_{0}|x_{1}\right)}\label{eq: S tot def}
\end{align}
\end{subequations}where we have denoted the composition of $\mathrm{R}$
and $\mathrm{T}$ as a composite operator $\mathrm{C}$, \emph{i.e.
$\mathbb{P}^{\mathrm{C}}=\left[\mathbb{P}^{\mathrm{R}}\right]^{\mathrm{T}}.$
}We note that the total entropy production satisfies additive property
when connecting two time intervals, \emph{i.e., }for $0<s<t$,
\begin{equation}
S_{\mathrm{tot}}(x_{0:t})=S_{\mathrm{tot}}(x_{0:s})+S_{\mathrm{tot}}(x_{s:t}).\label{eq: additivity of Stot}
\end{equation}

The composite operator $\mathrm{C}$ is generally not involutive \footnote{This can be easily seen by considering the case for only one time
step. We have $P_{0,1}^{\mathrm{C}}(x_{0},x_{1})=P_{1}(x_{1})P_{1|0}(x_{0}|x_{1})$
from Equation (\ref{eq: R and T composition}). Thus, we have $P_{0,1}^{\mathrm{CC}}(x_{0},x_{1})$
equal to $P_{1}^{\mathrm{C}}(x_{1})P_{1|0}^{\mathrm{C}}(x_{0}|x_{1})$.
With $P_{0,1}^{\mathrm{C}}$ given above, we get $P_{1}^{\mathrm{C}}=P_{1}$
and $P_{1|0}^{\mathrm{C}}=P_{1}(x_{0})P_{1|0}(x_{1}|x_{0})/P_{0}^{\mathrm{C}}(x_{1})$.
Therefore, $P_{0,1}^{\mathrm{CC}}(x_{0},x_{1})=P_{1}(x_{1})P_{1}(x_{0})P_{1|0}(x_{1}|x_{0})/P_{0}^{\mathrm{C}}(x_{1})$
which is not $P_{0,1}(x_{0},x_{1})$ unless all $P_{0}$, $P_{1}$,
and $P_{0}^{\mathrm{C}}$ are the invariant distribution of the transition
matrix $P_{1|0}$.}. Thus, $S_{\mathrm{tot}}$ does not admit TFT and the odd-parity,
sufficient condition for iDFT. For rDFT, we note 
\begin{align}
\bar{S}_{\mathrm{tot}}(\omega) & \coloneqq\ln\frac{\mathrm{d}\mathbb{P}^{\mathrm{R}}\ \thinspace}{\mathrm{d}\mathbb{P}^{\mathrm{RC}}}(\omega)=\ln\frac{\mathrm{d}\mathbb{P}^{\mathrm{R}}\ \ \ }{\mathrm{d}\mathbb{P}^{\mathrm{RRT}}}(\omega).\label{eq:-16}
\end{align}
and thus 
\begin{equation}
\bar{S}_{\mathrm{tot}}(r(\omega))=\ln\frac{\mathrm{d}\mathbb{P}^{\mathrm{RT}}\ \ \ }{\mathrm{d}\mathbb{P}^{\mathrm{RRTT}}}(\omega)=\ln\frac{\mathrm{d}\mathbb{P}^{\mathrm{RT}}}{\mathrm{d}\mathbb{P}^{\mathrm{RR}}}(\omega)\label{eq: rDFT for Stot}
\end{equation}
which becomes $-S_{\mathrm{tot}}(\omega)$ if $\mathrm{R}$ is involutive,
\emph{i.e. }$\mathbb{P}^{\mathrm{RR}}=\mathbb{P}.$ Hence, if $\mathrm{R}$
is involutive, $S_{\mathrm{tot}}$ admits rDFT. Recall that the requirement
for $\mathrm{R}$ to be involutive is the terminal distribution of
the protocol reversed process to recover the initial distribution
of the original process, \emph{i.e. }$P_{t}^{\mathrm{R}}=P_{0}$.
This condition was discussed in \citep{crooks_entropy_1999,seifert_entropy_2005,seifert_stochastic_2012}.

\subsection{Difference between $S_{\mathrm{T}}$ and $S_{\mathrm{tot}}$\label{subsec:Difference-between-ST and Stot}}

The physical meanings of the two EPs discussed above are clearly different.
For a given trajectory $\omega=x_{0:t}$, the dissipation function
$S_{\mathrm{T}}$ quantifies the probability difference between observing
the trajectory $\omega=x_{0:t}$ and the order reversal of it $r(\omega)=x_{t:0}$
in the original process. On the other hand, the total EP, $S_{\mathrm{tot}}$,
quantifies the probability difference between observing a trajectory
$x_{0:t}$ in the original process and observing the order-reversed
trajectory $x_{t:0}$ in the protocol-reversed process $\mathrm{R}$.

In time homogeneous processes where all $M_{n}=M$, $\forall n=1,2,\cdots,t$,
their difference gives another EP,\begin{subequations}
\begin{align}
D & \coloneqq S_{\mathrm{T}}-S_{\mathrm{tot}}=\ln\frac{\mathrm{d}\mathbb{P}^{\mathrm{RT}}}{\mathrm{d}\mathbb{P}^{\mathrm{T}}\ \thinspace}\label{eq:-2}\\
 & =\ln\frac{P_{t}(X_{t})M_{t}(X_{t-1}|X_{t})\cdots M_{1}(X_{0}|X_{1})}{P_{0}(X_{t})M_{1}(X_{t-1}|X_{t})\cdots M_{t}(X_{0}|X_{1})}\label{eq:-6}\\
 & =\ln\frac{P_{t}(X_{t})}{P_{0}(X_{t})}=\ln\frac{\mathrm{d}\mathbb{P}\ \ }{\mathrm{d}\mathbb{P}^{\mathrm{D}}}\label{eq: dP/dPD in homogeneous}
\end{align}
\end{subequations}where the time homogeneous assumption kicks in
to eliminate all the $M$s. The corresponding joint probability in
the measure $\mathbb{P}^{\mathrm{D}}$ would be 
\begin{equation}
P_{0:t}^{\mathrm{D}}(x_{0:t})=\frac{P_{0}(x_{t})}{P_{t}(x_{t})}P_{0:t}(x_{0:t}).\label{eq:-59}
\end{equation}
This implies that the expectation of the dissipation function is bigger
than the total EP, 
\begin{equation}
\mathbb{E}[D]\ge0\Leftrightarrow\mathbb{E}[S_{\mathrm{T}}]\ge\mathbb{E}[S_{\mathrm{tot}}],\label{eq: E ST > E Stot}
\end{equation}
for time homogeneous processes. By Jensen's inequality, we also know
the equality holds if and only if $P_{t}=P_{0}.$

In the time homogeneous cases, the expectation of $D$ is actually
the Kullback-Leibler divergence between the terminal distribution
$P_{t}$ and initial distribution $P_{0}$. Without time homogeneity,
$D$ is generally not an EP since $\frac{P_{0}(x_{t})M_{1}(x_{t-1}|x_{t})\cdots M_{t}(x_{0}|x_{1})}{P_{t}(x_{t})M_{t}(x_{t-1}|x_{t})\cdots M_{1}(x_{0}|x_{1})}P_{0:t}(x_{0:t})$
is not generally normalizable and thus not a joint probability. This
also indicates that the expectation of $D$ in time inhomogeneous
systems does not generally have a definite sign.

The infinitesimal time interval limit $t\rightarrow0$ can be taken
to consider the \emph{entropy} \emph{production} \emph{rate }(EPR)
\citep{ge_extended_2009,ge_physical_2010}. In such limit, we see
that $\mathbb{E}[D]$ is in $O(t^{2})$ whereas both $\mathbb{E}[S_{\mathrm{T}}]$
and $\mathbb{E}[S_{\mathrm{tot}}]$ are $e_{\mathrm{p}}t+O(t^{2})$
with the same entropy production rate, $e_{\mathrm{p}}$. This is
one of the reason why the contradistinction between $S_{\mathrm{T}}$
and $S_{\mathrm{tot}}$ was not clear in the past literature. See
Appendix \hyperref{[}{B}{{]}}{B} for derivation.

Recall that when connecting two time intervals, $S_{\mathrm{tot}}$
satisfies the additive property whereas $S_{\mathrm{T}}$ does not!
Thus, when one integrate the EPR of $S_{\mathrm{T}}$ and $S_{\mathrm{tot}}$
over time, the value one gets is $\mathbb{E}[S_{\mathrm{tot}}]$ not
$\mathbb{E}[S_{\mathrm{T}}]$. This resolves the seemly contradicting
results that the expectations of the two EPs are different for any
finite time interval but with the same rate in infinitesimal time
interval.

\subsection{Total Heat, Excess Heat and Housekeeping Heat Dissipation}

One of the most important breakthrough in nonequilibrium thermodynamics
is the discovery of the heat dissipation in nonequilibrium steady
state (NESS) and its statistical properties \citep{evans_equilibrium_1994,gallavotti_dynamical_1995,oono_steady_1998,kurchan_fluctuation_1998,lebowitz_gallavotticohen-type_1999,qian_nonequilibrium_2001,hatano_steady-state_2001,jiang_mathematical_2004,ge_transient_2007,ge_physical_2010}.
By the energy conservation, this amount is also the amount of energy
required to sustain the NESS, historically called the \emph{housekeeping}
\emph{heat $Q_{\mathrm{hk}}$ }and is conventionally chosen to be
positive as a heat \emph{dissipated} \emph{by} the system.

To understand the housekeeping heat with our Markov chain paradigm,
we consider a time inhomogeneous $t$-step Markov chain with a transition
matrix $M_{n}$. The total heat dissipation for a trajectory $\omega=x_{0:t}$
is given by the transition matrices,
\begin{equation}
Q(\omega)=\ln\prod_{n=1}^{t}\frac{M_{n}(x_{n}|x_{n-1})}{M_{n}(x_{n-1}|x_{n})},\label{eq: total heat}
\end{equation}
for general systems even non-detailed balanced \citep{crooks_entropy_1999,qian_nonequilibrium_2001}.
 Without detailed balance, the probability flux between two states
$i,j$ at NESS at time $n$ is nonzero, 
\begin{equation}
0\neq J_{n}(j|i)=\pi_{n}(i)M_{n}(j|i)-\pi_{n}(j)M_{n}(i|j).\label{eq:-22}
\end{equation}
This is physically originated from the non-conservative force that
sustains NESS \citep{seifert_stochastic_2012}. With the invariant
distribution $\pi_{n}$ in non-detailed balanced systems, we can use
the so called (fluctuating) \emph{nonequilibrium} \emph{potential
}$\Phi_{n}(x)$ based on the invariant distribution at each time step
\cite{seifert_stochastic_2012}, 
\begin{equation}
\Phi_{n}(x)=-\ln\pi_{n}(x),\label{eq: nonequilibrium energy}
\end{equation}
which would be the potential of mean force if one chooses the free
energy of the entire system (the system of interest and the environment)
as the zero potential energy reference point \citep{thompson_nonlinear_2016}.
It is worth noting that for time homogeneous diffusion processes in
the thermodynamic limit, this \emph{definition} of the nonequilibrium
potential, with proper scaling, gives us a Lyapunov function of the
emerged, dissipative deterministic dynamics \cite{qian_kinematic_2017}.

The changes of the nonequilibrium energy due to a transition in microstate
would then be \emph{the} \emph{excess} \emph{heat} \emph{dissipation},\begin{subequations}
\begin{align}
Q_{\mathrm{ex}}(x_{0:t}) & =\sum_{n=1}^{t}\Phi_{n}(x_{n-1})-\Phi_{n}(x_{n})\\
 & =\ln\prod_{n=1}^{t}\frac{\pi_{n}(x_{n})}{\pi_{n}(x_{n-1})}.\label{eq:excess heat}
\end{align}
\end{subequations} The housekeeping heat (dissipation) is given by
the difference between the total heat dissipation $Q$ and the excess
heat dissipation $Q_{\mathrm{ex}}$, 
\begin{equation}
Q_{\mathrm{hk}}=Q-Q_{\mathrm{ex}}\label{eq:-23}
\end{equation}
This will become\begin{subequations}
\begin{align}
Q_{\mathrm{hk}}(\omega) & =\sum_{n=1}^{t}\ln\frac{M_{n}(X_{n}|X_{n-1})}{M_{n}^{\dagger}(X_{n}|X_{n-1})}\label{eq:Qhk M/M}\\
 & =\ln\frac{P_{0}(X_{0})}{P_{0}(X_{0})}\prod_{n=1}^{t}\frac{M_{n}(X_{n}|X_{n-1})}{M_{n}^{\dagger}(X_{n}|X_{n-1})}\label{eq:-61}\\
 & =\ln\frac{\mathrm{d}\mathbb{P}\ }{\mathrm{d}\mathbb{P}^{\dagger}}(\omega)\label{eq: housekeeping heat}
\end{align}
\end{subequations}which shows that it is an EP with corresponding
the CPM operator $\dagger$, thus has a non-negative expectation,
and admits both IFT and GFR. If the system possesses detailed balance,
then $Q_{\mathrm{hk}}=0$. Straight from the definition in Equation
(\ref{eq:Qhk M/M}) that, similar to $S_{\mathrm{tot}}$, the housekeeping
heat is also additive when connecting time intervals,
\begin{equation}
Q_{\mathrm{hk}}(x_{0:t})=Q_{\mathrm{hk}}(x_{0:s})+Q_{\mathrm{hk}}(x_{s:t}).\label{eq: additivitiy of Qhk}
\end{equation}
With detailed balance, $Q_{\mathrm{hk}}=0$ and the excess heat dissipation
$Q_{\mathrm{ex}}$ reduces to the total heat dissipation $Q$ \citep{hatano_steady-state_2001}.

Since $\dagger$ is involutive, the housekeeping heat $Q_{\mathrm{hk}}$
generally admits iDFT. To show whether it admits rDFT or not, we compute
the housekeeping heat in the protocol-reversed process evaluated at
the order-reversed trajectory,\begin{subequations}
\begin{align}
\bar{Q}_{\mathrm{hk}}(r(\omega)) & =\ln\frac{\mathrm{d}\mathbb{P}^{\mathrm{RT}}\ }{\mathrm{d}\mathbb{P}^{\mathrm{R}\dagger\mathrm{T}}}(\omega)\label{eq:-24}\\
 & =\ln\prod_{n=1}^{t}\frac{M_{t+1-n}\left(X_{t-n}|X_{t+1-n}\right)}{M_{t+1-n}^{\dagger}\left(X_{t-n}|X_{t+1-n}\right)}\label{eq:-62}\\
 & =\ln\prod_{n=1}^{t}\frac{M_{t+1-n}^{\dagger}\left(X_{t+1-n}|X_{t-n}\right)}{M_{t+1-n}\left(X_{t+1-n}|X_{t-n}\right)}\\
 & =-Q_{\mathrm{hk}}(\omega).
\end{align}
\end{subequations}where we have used $\frac{M_{k}\left(j|i\right)}{M_{k}^{\dagger}\left(j|i\right)}=\frac{M_{k}^{\dagger}\left(i|j\right)}{M_{k}\left(i|j\right)}$.
Hence, the housekeeping heat $Q_{\mathrm{hk}}(\omega)$ generally
admits rDFT.

Lastly, for time homogeneous processes, the housekeeping heat is the
dissipation function $S_{\mathrm{T}}$ starting at the invariant distribution
$\pi$ since, as we have discussed, the dual process is equivalent
to the time reversal if the system reaches the invariant state. Since
$S_{\mathrm{T}}$ admits TFT for arbitrary initial distribution, this
also means that $Q_{\mathrm{hk}}$ also admits TFT for time homogeneous
processes starting at the invariant steady state \citep{gallavotti_dynamical_1995,kurchan_fluctuation_1998,lebowitz_gallavotticohen-type_1999,jiang_mathematical_2004,ge_transient_2007}.

\subsection{Exergy, Excess Work, and the Non-adiabatic Entropy Production}

With the notion of the nonequilibrium energy $\Phi_{n}$ defined
in Equation (\ref{eq: nonequilibrium energy}) at each step for general
non-detailed balance systems, the \emph{excess work }done \emph{by
}the\emph{ }system for a trajectory $x_{0:t}$ is then the difference
between the nonequilibrium potential dissipation $-\Delta\Phi$ and
the excess heat dissipated $Q_{\mathrm{ex}}$ as illustrated in Figure
\ref{fig:Discrete-time-heat-and-work},
\begin{subequations}
\begin{align}
W_{\mathrm{ex}}(x_{0:t}) & =-\Delta\Phi(x_{0:t})-Q_{\mathrm{ex}}(x_{0:t})\\
 & =\ln\prod_{n=0}^{t-1}\frac{\pi_{n+1}(x_{n})}{\pi_{n}(x_{n})},\label{eq: excess work}
\end{align}
which is the change in the nonequilibrium potential due to change
in the transition matrices (and thus the corresponding invariant distribution).
\begin{figure}
\begin{centering}
\includegraphics[width=1\columnwidth]{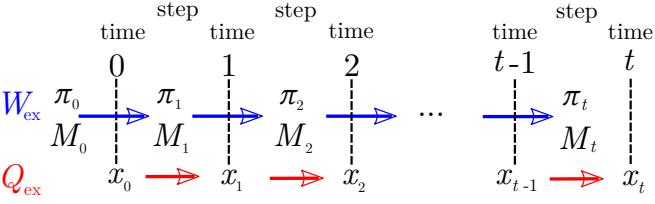}
\par\end{centering}
\caption{Excess Heat and excess work definition in a discrete time Markov Chain.
$x_{n}$ are the state of the system at time $n$. $M_{n}$ are the
transition matrices at the $n$th time step and $\pi_{n}$ are the
corresponding unique invariant distribution. \label{fig:Discrete-time-heat-and-work}}
\end{figure}
\end{subequations}

The fluctuating relative entropy between $P_{n}$ and $\pi_{n}$ is
given by
\begin{equation}
F_{n}(x)=\Phi_{n}(x)-S_{n}(x)=\ln\frac{P_{n}(x)}{\pi_{n}(x)}.\label{eq: NEF for NDB}
\end{equation}
For systems with detailed balance, the sum of this fluctuating relative
entropy and the free energy defined in classical equilibrium thermodynamics
was called \emph{nonequilibrium free energy} in \cite{parrondo_thermodynamics_2015}.
For this reason, the relative entropy also got a name \emph{nonsteady-state
addition }(to free energy) in \cite{riechers_fluctuations_2017}.
Furthermore, it was shown in \cite{qian_relative_2001} that this
relative entropy itself could be understood physically as a ``free
energy'' as well. To avoid possible confusion on the terminology
in this paper, we would follow \cite{altaner_nonequilibrium_2017}
and call this (fluctuating) \emph{exergy} from now on.. For a trajectory
$\omega=x_{0:t}$, the exergy that got absorbed by the system is then
\begin{equation}
\Delta F(x_{0:t})=\ln\frac{P_{t}(x_{t})}{\pi_{t}(x_{t})}-\ln\frac{P_{0}(x_{0})}{\pi_{0}(x_{0})}.\label{eq: Free Energy difference}
\end{equation}

The difference between the exergy dissipation $-\Delta F$ and the
excess work done \emph{by }the system $W_{\mathrm{ex}}$ then gives
us the \emph{non-adiabatic} EP\emph{,} $S_{\mathrm{na}}$, defined
in \citep{esposito_three_2010},\begin{subequations}
\begin{align}
S_{\mathrm{na}}(\omega) & =-\Delta F(\omega)-W_{\mathrm{ex}}(\omega)\label{eq:-74}\\
 & =Q_{\mathrm{ex}}(\omega)+\Delta S(\omega)\\
 & =\ln\frac{\mathrm{d}\mathbb{P}\ \ \ \ \thinspace}{\mathrm{d}\mathbb{P}^{\mathrm{R}\dagger\mathrm{T}}}(\omega).\label{eq: non adiabatic ep}
\end{align}
\end{subequations}The equivalence between the last two lines can
be seen by a direct computation of $P_{0:t}^{\mathrm{R}\dagger\mathrm{T}}(x_{0:t})$.
We note that $S_{\mathrm{na}}$ becomes the exergy dissipation $-\Delta F$
in time homogeneous processes since $W_{\mathrm{ex}}=0$ in time homogeneous
processes. We also note that the non-adiabatic EP is also additive
when connecting time interval,
\begin{equation}
S_{\mathrm{na}}(x_{0:t})=S_{\mathrm{na}}(x_{0:s})+S_{\mathrm{na}}(x_{s:t})\label{eq: Sna additive}
\end{equation}
 which is obvious from its relation to $W_{\mathrm{ex}}$ and $\Delta F$.
Note that $S_{\mathrm{na}}$ reduces to the dissipative work $W_{\mathrm{d}}$
defined in \citep{jarzynski_nonequilibrium_1997,crooks_nonequilibrium_1998}
for systems with detailed balance.

Similar to the total EP $S_{\mathrm{tot}}$, the non-adiabatic EP
$S_{\mathrm{na}}$ admit neither the TFT nor the odd-parity, sufficient
condition for iDFT since the composite operator $\mathrm{R\dagger T}$
is not involutive in general. For rDFT, we compute\begin{subequations}
\begin{align}
\bar{S}_{\mathrm{na}}(r(\omega)) & =\ln\frac{\mathrm{d}\mathbb{P}^{\mathrm{RT}}\ }{\mathrm{d}\mathbb{P}^{\mathrm{RR}\dagger}}(\omega)\label{eq:-80}\\
 & =\ln\frac{P_{t}(X_{t})}{P_{t}^{\mathrm{R}}(X_{0})}\prod_{n=1}^{t}\frac{M_{n}(X_{n-1}|X_{n})}{M_{n}^{\dagger}(X_{n}|X_{n-1})}\label{eq:-30}
\end{align}
\end{subequations}where the denominator can be obtained by using
Equation (\ref{eq: involutive R}) for $\mathbb{P}^{\mathrm{RR}}$
and apply $\dagger$ on it. Now, since 
\begin{equation}
\frac{M_{n}(X_{n-1}|X_{n})}{M_{n}^{\dagger}(X_{n}|X_{n-1})}=\frac{\pi_{n}(X_{n-1})}{\pi_{n}(X_{n})},\label{eq:-31}
\end{equation}
we get 
\begin{equation}
\bar{S}_{\mathrm{na}}(r(\omega))=\ln\frac{P_{0}(X_{0})}{P_{t}^{\mathrm{R}}(X_{0})}-S_{\mathrm{na}}(\omega).\label{eq: P0 =00003D Pt^R requirement for Sna}
\end{equation}
We see that the condition for the odd parity to hold, $\bar{S}_{\mathrm{na}}(r(\omega))=-S_{\mathrm{na}}(\omega)$,
is $P_{0}=P_{t}^{\mathrm{R}},$ \emph{i.e.}, $\mathrm{R}$ to be involutive!
Hence, similar to the total EP $S_{\mathrm{tot}}$, the non-adiabatic
EP $S_{\mathrm{na}}$ admits rDFT if the CPM operator $\mathrm{R}$
is involutive. The rDFT of $S_{\mathrm{na}}$ is an extension to Crooks'
fluctuation theorem \citep{crooks_nonequilibrium_1998,crooks_entropy_1999,crooks_path-ensemble_2000}.

\subsection{Martingale Properties of Entropy Productions}

With our measure-theoretic understanding of EPs, more statistical
properties of EPs could be found by revealing more on the mathematical
properties of their corresponding RNDs and CPM operators. For example,
by recognizing the RND of $Q_{\mathrm{hk}}$, $\exp(-Q_{\mathrm{hk}})$,
is a martingale, a statistics of the infimum of $Q_{\mathrm{hk}}$
is introduced in \cite{chetrite_two_2011,neri_statistics_2017,chetrite_martingale_2019}.
Here, we shall discuss the martingale properties of the four EPs and
the conditions for the exponential of the negative of them to be a
martingale.

In our discrete time Markov chain paradigm, a functional $M(X_{0:t})$
is a martingale if it satisfies 
\begin{equation}
\mathbb{E}[M(X_{0:t})|X_{0:s}]=M(X_{0:s})\label{eq: martingale def}
\end{equation}
$\forall s\in\left\{ 0,1,\cdots,t\right\} $. It can be shown rather
straightforwardly that since $S_{\mathrm{T}}$ is not additive in
time, $\exp(-S_{\mathrm{T}})$ would not generally be a martingale.
Thus, in the following discussion, we will focus on $S_{\mathrm{tot}}$,
$Q_{\mathrm{hk}}$, and $S_{\mathrm{na}}$.

We note that all of $S_{\mathrm{tot}}$, $Q_{\mathrm{hk}}$, and $S_{\mathrm{na}}$
are additive in time. Therefore, for $\exp(-S_{\nu})$ to be a martingale
where $\nu=\mathrm{RT},\dagger,$or $\mathrm{R\dagger\mathrm{T}}$,
we need\begin{subequations}
\begin{align}
\mathbb{E}[e^{-S_{\nu}(X_{0:t})}|X_{0:s}] & =e^{-S_{\nu}(X_{0:s})}\mathbb{E}[e^{-S_{\nu}(X_{s:t})}|X_{0:s}]\label{eq: additivitiy}\\
 & =e^{-S_{\nu}(X_{0:s})}\mathbb{E}[e^{-S_{\nu}(X_{s:t})}|X_{s}]\label{eq: Markovian}\\
 & \ensuremath{\stackrel{!}{=}}e^{-S_{\nu}(X_{0:s})}.\label{eq: reduced relation WTS}
\end{align}
\end{subequations}We thus want to have
\begin{equation}
\mathbb{E}[e^{-S_{\nu}(X_{s:t})}|X_{s}]=1,\label{eq: actual reduced relation WTS for martingale}
\end{equation}
$\forall s\in\left\{ 0,1,\cdots,t\right\} $.

By using the definition for $\nu=\mathrm{RT},\dagger$, and $\mathrm{R}\dagger\mathrm{T}$,
one will find \begin{subequations}
\begin{align}
\mathbb{E}[e^{-S_{\mathrm{tot}}(X_{s:t})}|X_{s}] & =\frac{P_{\mathrm{end}}^{\mathrm{R}}(X_{s})}{P_{s}(X_{s})},\label{eq: Stot conditional exp}\\
\mathbb{E}[e^{-Q_{\mathrm{hk}}(X_{s:t})}|X_{s}] & =1,\text{ and }\label{eq: Qhk conditional exp}\\
\mathbb{E}[e^{-S_{\mathrm{na}}(X_{s:t})}|X_{s}] & =\frac{P_{\mathrm{end}}^{\mathrm{R}\dagger}(X_{s})}{P_{s}(X_{s})}\label{eq: Sna conditional exp}
\end{align}
\end{subequations}where 
\begin{align}
P_{\mathrm{end}}^{\mathrm{R}}(x_{s}) & =\sum_{x_{s+1:t}}P_{t}(x_{t})\prod_{n=0}^{t-s-1}M_{t-n}(x_{t-n-1}|x_{t-n})\text{ and }\nonumber \\
P_{\mathrm{end}}^{\mathrm{R\dagger}}(x_{s}) & =\sum_{x_{s+1:t}}P_{t}(x_{t})\prod_{n=0}^{t-s-1}M_{t-n}^{\dagger}(x_{t-n-1}|x_{t-n})\label{eq: P end for R and Rdagger}
\end{align}
are the terminal distributions of the processes $\mathrm{R}$ and
$\mathrm{R}\dagger$ defined on the time interval $s:t$.

This shows that $\exp(-Q_{\mathrm{hk}})$ is always a martingale which
implies that $Q_{\mathrm{hk}}$ is a submartingale satisfying 
\begin{equation}
\mathbb{E}[Q_{\mathrm{hk}}(X_{0:t})|X_{0:s}]\ge Q_{\mathrm{hk}}(X_{0:s})\label{eq: super martingale}
\end{equation}
by the convexity of negative logarithm. Since $s$ is arbitrary in
$0:t$, the RHS of Equation (\ref{eq: Stot conditional exp}) and
(\ref{eq: Sna conditional exp}) needs to be 1 for all $x_{s}$ and
$s$. This means that $\exp(-S_{\mathrm{tot}})$ or $\exp(-S_{\mathrm{na}})$
are only martingale when the reversal $\mathrm{R}$ or $\mathrm{R\dagger}$
recovers all the marginal distributions in a reversed order: $P_{t}\rightarrow P_{t-1}\rightarrow\cdots\rightarrow P_{0}$
in the reversed process, which is not generally true. One exception
is when the dynamics is time homogeneous and also starts with the
invariant distribution such that $P_{n}=\pi$, $\forall n\in\left\{ 0,1,\cdots,t\right\} $.

\subsection{Summary}

The properties of the four EPs in physics and chemistry including
various FRs we have discussed above have been summarized in Table
\ref{tab:Summary-of-the}. Here, we note that the three reversals
$\mathrm{T}$, $\mathrm{R}$, and $\dagger$ are actually related.
By direct computation, one can get 
\begin{equation}
\frac{\mathrm{d}\mathbb{P}\ \ \ }{\mathrm{d}\mathbb{P}^{\mathrm{RT}}}=\frac{\mathrm{d}\mathbb{P}\ }{\mathrm{d}\mathbb{P}^{\dagger}}\frac{\mathrm{d}\mathbb{P}\ \ \ \ \thinspace}{\mathrm{d}\mathbb{P}^{\mathrm{R}\dagger\mathrm{T}}}\label{eq: tri-relation between reversals}
\end{equation}
which leads to the famous decomposition of the total EP introduced
in \citep{ge_extended_2009},
\begin{equation}
S_{\mathrm{tot}}=S_{\mathrm{na}}+Q_{\mathrm{hk}}.\label{eq:-33}
\end{equation}
Since $Q_{\mathrm{hk}}=0$ when the system possesses detailed balance,
we also see that $S_{\mathrm{tot}}\equiv S_{\mathrm{na}}$ in detailed
balance systems.

\subsection{Two Fluctuation Relations for Heat and Work}

As an another demonstration on how the formalism can help us obtain
statistical properties of EP-related quantities, we note that there
is another generally valid FR called \emph{differential fluctuation
theorem }for the work done \emph{by} the system $W$, derived in \citep{jarzynski_hamiltonian_2000,maragakis_differential_2008}
and experimental verified in \citep{hoang_experimental_2018} for
detailed balanced systems. Here we provide a more general derivation
to extend it to non-detailed balanced systems. The key observation
is that the excess work defined in Equation (\ref{eq: excess work})
always has odd parity under the composite CPM operator $\mathrm{C}\equiv\mathrm{RT}$,
\emph{i.e}. $\bar{W}_{\mathrm{ex}}(r(\omega))=-W_{\mathrm{ex}}(\omega)$.
Knowing this, we can consider the joint probability of the work $W_{\mathrm{ex}}$,
initial state $X_{0}$, and the terminal state $X_{t}$,
\begin{align}
P_{W_{\mathrm{ex}},X_{0},X_{t}}(w,x_{0},x_{t}) & =\mathbb{E}[\mathbb{I}_{\{W_{\mathrm{ex}}\in\mathrm{d}w,X_{0}=x_{0},X_{t}=x_{t}\}}].\label{eq:-34}
\end{align}

The joint probability under the measure $\mathbb{P}^{\mathrm{R}\dagger\mathrm{T}}$
is then given by\begin{subequations}
\begin{align}
 & P_{W_{\mathrm{ex}},X_{0},X_{t}}^{\mathrm{R}\dagger\mathrm{T}}(w,x_{0},x_{t})\nonumber \\
 & =\mathbb{E}[e^{-S_{\mathrm{na}}}\mathbb{I}_{\{W_{\mathrm{ex}}\in\mathrm{d}w,X_{0}=x_{0},X_{t}=x_{t}\}}]\\
 & =e^{w+\Delta F(x_{0},x_{t})}\mathbb{E}[\mathbb{I}_{\{W_{\mathrm{ex}}\in\mathrm{d}w,X_{0}=x_{0},X_{t}=x_{t}\}}]\label{eq:-35}\\
 & =e^{w+\Delta F(x_{0},x_{t})}P_{W_{\mathrm{ex}},X_{0},X_{t}}(w,x_{0},x_{t})
\end{align}
\end{subequations}where we have used the fact that the exergy increment
$\Delta F(\omega)=\ln\frac{P_{t}(X_{t})}{\pi_{t}(X_{t})}-\ln\frac{P_{0}(X_{0})}{\pi_{0}(X_{0})}$
is a function of $X_{0}$ and $X_{t}$.

By using $\bar{W}_{\mathrm{ex}}(r(\omega))=-W_{\mathrm{ex}}(\omega)$,
we thus have
\begin{equation}
\frac{P_{W_{\mathrm{ex}},X_{0},X_{t}}(w,x_{0},x_{t})}{P_{\bar{W}_{\mathrm{ex}},X_{0}^{\mathrm{T}},X_{t}^{\mathrm{T}}}^{\mathrm{R}\dagger}(-w,x_{t},x_{0})}=e^{-\Delta F(x_{0},x_{t})-w}.\label{eq:-36}
\end{equation}
Also, a similar differential fluctuation theorem for the excess heat
dissipated $Q_{\mathrm{ex}}(\omega)$ can also be derived, 
\begin{equation}
\frac{P_{Q_{\mathrm{ex}},X_{0},X_{t}}(q,x_{0},x_{t})}{P_{\bar{Q}_{\mathrm{ex}},X_{0}^{\mathrm{T}},X_{t}^{\mathrm{T}}}^{\mathrm{R}\dagger}(-q,x_{t},x_{0})}=e^{q+\Delta S(x_{0},x_{t})}\label{eq:-37}
\end{equation}
since $\bar{Q}_{\mathrm{ex}}(r(\omega))=-Q_{\mathrm{ex}}(\omega)$
and $S_{\mathrm{na}}=-\Delta F-W_{\mathrm{ex}}=Q_{\mathrm{ex}}+\Delta S$.

\section{Entropy Productions in Constant-noise Diffusion Processes\label{sec:Entropy-Productions-in Diffusion}}

We will now briefly go through how to use measure-theoretic probability
theory and the CPM operator formalism to derive the stochastic integral
formulas for the four EPs in time inhomogeneous constant-noise diffusion
processes. With the formulas, one can further derive the expression
for the moments of EPs with Ito calculus. In a constant-noise diffusion
process in $\mathbb{R}^{n}$, the probability density of the microstate
variable $\mathbf{X}_{t}$, $p(\mathbf{x},t)$, is governed by the
Fokker-Planck Equation, 
\begin{equation}
\frac{\partial}{\partial t}p(\mathbf{x},t)=-\nabla\cdot\mathbf{J}[p(\mathbf{x},t)]\label{eq: FPE}
\end{equation}
with probability flux given by 
\begin{equation}
\mathbf{J}[p(\mathbf{x},t)]=\mathbf{b}(\mathbf{x},t)p(\mathbf{x},t)-\mathbf{D}\nabla p(\mathbf{x},t).\label{eq: J}
\end{equation}
and the stochastic trajectory governed by the stochastic differential
equation, 
\begin{equation}
\mathrm{d}\mathbf{X}_{t}=\mathbf{b}(\mathbf{X}_{t},t)\mathrm{d}t+\mathbf{\Gamma}\mathrm{d}\mathbf{W}_{t},\label{eq: SDE}
\end{equation}
where $\mathbf{D}=\mathbf{\Gamma}\mathbf{\Gamma}^{\mathsf{T}}/2$
is a constant diffusion matrix ($\mathsf{T}$ denoting transpose)
and $\mathbf{W}_{t}$ is the $n$ dimensional Wiener processes (Brownian
motions) with each component independent and having unit strength
of noise.

To derive the stochastic integral formula for the four EPs from their
RND definitions, we rely on Girsanov theorem \cite{jiang_mathematical_2004}
to give us the RND to ``kill'' the drift $\mathbf{b}(\mathbf{X}_{t},t)$
in the dynamics of $\mathbf{X}_{t}$. Given a time interval $t\in\left[0,\tau\right]$
and the RND\begin{subequations}
\begin{align}
\frac{\mathrm{d}\mathbb{P}^{\mathrm{B}}}{\mathrm{d}\mathbb{P}_{\ \thinspace}} & =e^{-\frac{1}{2}\int_{0}^{\tau}\mathbf{D}^{-1}\mathbf{b}\cdot\mathrm{d}\mathbf{X}_{t}+\frac{1}{4}\int_{0}^{\tau}\mathbf{b}\cdot\mathbf{D}^{-1}\mathbf{b}\mathrm{d}t}\label{eq: GSN Ito}\\
 & =e^{-\frac{1}{2}\int_{0}^{\tau}\mathbf{D}^{-1}\mathbf{b}\circ\mathrm{d}\mathbf{X}_{t}+\frac{1}{2}\int_{0}^{\tau}\nabla\cdot\mathbf{b}\mathrm{d}t+\frac{1}{4}\int_{0}^{\tau}\mathbf{b}\cdot\mathbf{D}^{-1}\mathbf{b}\mathrm{d}t},\label{eq:GSN Strat}
\end{align}
\end{subequations} the probability density of the process $\mathbf{X}_{t}$
under the measure $\mathbb{P}^{\mathrm{B}}$ satisfies $\frac{\partial}{\partial t}p^{\mathrm{B}}(\mathbf{x},t)=\nabla\cdot\mathbf{D}\nabla p^{\mathrm{B}}(\mathbf{x},t)$,
\emph{i.e.} $\mathbf{X}_{t}$ is a Brownian motion with strength $\mathbf{\Gamma}$
under $\mathbb{P}^{\mathrm{B}}$. The CPM operator B ``kills'' the
drift by changing of the probability measure from $\mathbb{P}$ to
$\mathbb{P}^{\mathrm{B}}$. We note that the first stochastic integral
in Equation (\ref{eq: GSN Ito}) is an Ito integral and the first
stochastic integral in Equation (\ref{eq:GSN Strat}) is a Stratonovich
integral. We rewrote the stochastic integral into a Stratonovich form
for later convenience when considering time reversals.

\subsection{Dissipation Function}

To use Equation (\ref{eq:GSN Strat}) to get $S_{\mathrm{T}}$, we
perform the decomposition of its corresponding RND,
\begin{equation}
\frac{\mathrm{d}\mathbb{P}\ \thinspace}{\mathrm{d}\mathbb{P}^{\mathrm{T}}}=\frac{\mathrm{d}\mathbb{P}\ \ }{\mathrm{d}\mathbb{P}^{\mathrm{B}}}\frac{\mathrm{d}\mathbb{P}^{\mathrm{B}}\ \thinspace}{\mathrm{d}\mathbb{P}^{\mathrm{BT}}}\frac{\mathrm{d}\mathbb{P}^{\mathrm{BT}}}{\mathrm{d}\mathbb{P}^{\mathrm{T}}\ \thinspace}.\label{eq: decomposing dP/dP^T}
\end{equation}
We already have the first term $\frac{\mathrm{d}\mathbb{P}\ \ }{\mathrm{d}\mathbb{P}^{\mathrm{B}}}=-\frac{\mathrm{d}\mathbb{P}^{\mathrm{B}}}{\mathrm{d}\mathbb{P}_{\ \thinspace}}$
by Equation (\ref{eq:GSN Strat}). For the third term, it can be rewritten
as
\begin{align}
\frac{\mathrm{d}\mathbb{P}^{\mathrm{B}}}{\mathrm{d}\mathbb{P}_{\ \thinspace}}(r(\omega))= & \exp\{\frac{1}{2}\int_{0}^{\tau}\mathbf{D}^{-1}\mathbf{b}(\mathbf{X}_{t},\tau-t)\circ\mathrm{d}\mathbf{X}_{t}\nonumber \\
 & +\frac{1}{2}\int_{0}^{\tau}\nabla\cdot\mathbf{b}(\mathbf{X}_{t},\tau-t)\mathrm{d}t\nonumber \\
 & +\frac{1}{4}\int_{0}^{\tau}\mathbf{b}\cdot\mathbf{D}^{-1}\mathbf{b}(\mathbf{X}_{t},\tau-t)\mathrm{d}t\}\label{eq:d PBT / dPT}
\end{align}
where I have used the change of integration variable for the three
integrals in the exponent using
\begin{align}
\int_{0}^{\tau}\mathbf{b}(\mathbf{X}_{\tau-t},t)\circ\mathrm{d}\mathbf{X}_{\tau-t} & =-\int_{0}^{\tau}\mathbf{b}(\mathbf{X}_{t},\tau-t)\circ\mathrm{d}\mathbf{X}_{t}\text{ and }\nonumber \\
\int_{0}^{\tau}f(\mathbf{X}_{\tau-t},t)\mathrm{d}t & =\int_{0}^{\tau}f(\mathbf{X}_{t},\tau-t)\mathrm{d}t.\label{eq: b o dX sub t}
\end{align}
We note that the first equality in Equation (\ref{eq: b o dX sub t})
is true since it is a Stratonovich integral. If the stochastic integral
is in any other integration scheme, the change of integration variable
will lead to a change of integral scheme.

The second term in Equation (\ref{eq: decomposing dP/dP^T}) is the
RND of the dissipation function in Brownian motion. Since $\mathbf{X}_{t}$
is drift-free under $\mathbb{P}^{\mathrm{B}}$ and the noise strength
is a constant, the conditional probability for having a path $\omega=x_{0:t}$
conditioning that it starts at $x_{0}$ is the same as the one for
the reversed path $r(\omega)=x_{t:0}$ conditioning that it starts
at $x_{t}$ due to the symmetry of Brownian motion. We thus see that
the $\mathrm{T}$ operator only changes $\mathbb{P}^{\mathrm{B}}$
by the initial probability density $p_{0}$ from evaluating at $\mathbf{X}_{0}$,
$p_{0}(\mathbf{X}_{0})$, to evaluating at $\mathbf{X}_{\tau}$, $p_{0}(\mathbf{X}_{\tau})$,
and the change of measure is completed by
\begin{equation}
\frac{\mathrm{d}\mathbb{P}^{\mathrm{B}}\ \thinspace}{\mathrm{d}\mathbb{P}^{\mathrm{BT}}}=\frac{p_{0}(\mathbf{X}_{0})}{p_{0}(\mathbf{X}_{\tau})}\label{eq: dP0/dP0^T}
\end{equation}
where I have also used that the operator B leaves the initial probability
density invariant, $p_{0}^{\mathrm{B}}=p_{0}$.

Putting Equation (\ref{eq: decomposing dP/dP^T}-\ref{eq: dP0/dP0^T})
together, we arrive the dissipation function for a time inhomogeneous
constant-noise diffusion:
\begin{align}
S_{\mathrm{T}}= & \ln\frac{p_{0}(\mathbf{X}_{0})}{p_{0}(\mathbf{X}_{\tau})}+\int_{0}^{\tau}\mathbf{D}^{-1}\bar{\mathbf{b}}\circ\mathrm{d}\mathbf{X}_{t}\nonumber \\
 & +\frac{1}{2}\int_{0}^{\tau}\delta\left(\nabla\cdot\mathbf{b}\right)\mathrm{d}t+\frac{1}{4}\int_{0}^{\tau}\delta\left(\mathbf{b}\cdot\mathbf{D}^{-1}\mathbf{b}\right)\mathrm{d}t\label{eq: ST time inhomogeneous}
\end{align}
where we have used the notation\begin{subequations}
\begin{align}
\bar{\mathbf{b}}(\mathbf{X}_{t},t) & =\left[\mathbf{b}(\mathbf{X}_{t},t)+\mathbf{b}(\mathbf{X}_{t},\tau-t)\right]/2,\text{ and}\label{eq:bbar}\\
\delta f(\mathbf{X}_{t},t) & =f(\mathbf{X}_{t},\tau-t)-f(\mathbf{X}_{t},t).\label{eq:delta b}
\end{align}
\end{subequations} The stochastic integral expression of the dissipation
function $S_{\mathrm{T}}$ in Equation (\ref{eq: ST time inhomogeneous})
in general time inhomogeneous constant-noise diffusion process, as
far as we know, is a new result.

We note that if the drift is \emph{symmetric in the interval} $t\in\left[0,\tau\right]$,
\emph{i.e.} $\mathbf{b}(\cdot,t)=\mathbf{b}(\cdot,\tau-t)$, then
the two terms with $\delta$ become zero. The expression for dissipation
function then reduces to
\begin{equation}
S_{\mathrm{T}}=\ln\frac{p_{0}(\mathbf{X}_{0})}{p_{0}(\mathbf{X}_{\tau})}+Q\label{eq: ST for time symmetric b}
\end{equation}
where $Q=\int_{0}^{\tau}\mathbf{D}^{-1}\mathbf{b}\circ\mathrm{d}\mathbf{X}_{t}$
is the heat dissipation. As time homogeneous process being symmetric
in the interval, this is consistent with \cite{seifert_stochastic_2012}.
It can also be seen from this expression that, consistent with our
understanding in our Markov chain paradigm, $S_{\mathrm{T}}$ is not
generally additive when connecting time intervals.

\subsection{Total Entropy Production}

With $S_{\mathrm{tot}}=\ln\frac{\mathrm{d}\mathbb{P}\ \ \ }{\mathrm{d}\mathbb{P}^{\mathrm{RT}}}$,
we derive the formula of it in a similar decomposition of the corresponding
RND,
\begin{equation}
\frac{\mathrm{d}\mathbb{P}\ \ \ }{\mathrm{d}\mathbb{P}^{\mathrm{RT}}}(\omega)=\frac{\mathrm{d}\mathbb{P}\ \ }{\mathrm{d}\mathbb{P}^{\mathrm{B}}}\frac{\mathrm{d}\mathbb{P}^{\mathrm{B}}\ \ \ }{\mathrm{d}\mathbb{P}^{\mathrm{BRT}}}\frac{\mathrm{d}\mathbb{P}^{\mathrm{BRT}}}{\mathrm{d}\mathbb{P}^{\mathrm{RBT}}}\frac{\mathrm{d}\mathbb{P}^{\mathrm{RBT}}}{\mathrm{d}\mathbb{P}^{\mathrm{RT}}\ \thinspace}.\label{eq: decompoing dP/dP^RT}
\end{equation}
We can find the second and the fourth RNDs in a similar way. One has
\begin{align}
\frac{\mathrm{d}\mathbb{P}^{\mathrm{B}}\ \ \ }{\mathrm{d}\mathbb{P}^{\mathrm{BRT}}} & =\frac{p_{0}(\mathbf{X}_{0})}{p_{\tau}(\mathbf{X}_{\tau})}\label{eq: dP0/dP0^RT}
\end{align}
since $\mathrm{RT}$ only changes the initial distribution for $\mathbb{P}^{\mathrm{B}}$
and
\begin{align}
\frac{\mathrm{d}\mathbb{P}^{\mathrm{RBT}}}{\mathrm{d}\mathbb{P}^{\mathrm{RT}}\ \thinspace}(\omega)= & \exp\{\frac{1}{2}\int_{0}^{\tau}\mathbf{D}^{-1}\mathbf{b}(\mathbf{X}_{t},t)\circ\mathrm{d}\mathbf{X}_{t}\nonumber \\
 & +\frac{1}{2}\int_{0}^{\tau}\nabla\cdot\mathbf{b}(\mathbf{X}_{t},t)\mathrm{d}t\nonumber \\
 & +\frac{1}{4}\int_{0}^{\tau}\mathbf{b}\cdot\mathbf{D}^{-1}\mathbf{b}(\mathbf{X}_{t},t)\mathrm{d}t\}.\label{eq: dP0^RT/dP^RT}
\end{align}
by replacing $\mathbf{b}(\cdot,t)$ with $\mathbf{b}^{\mathrm{R}}(\cdot,t)=\mathbf{b}(\cdot,\tau-t)$
in Equation (\ref{eq:d PBT / dPT}). 

Now, for the third term in Equation (\ref{eq: decompoing dP/dP^RT}),
since B kills the drift for both $\mathbf{b}$ and $\mathbf{b}^{\mathrm{R}}$,
the effect of R after composed with B, regardless of the order, is
just a change in the initial distribution. BR is thus the same as
RB since B doesn't change the initial distribution. This means the
operator B commutes with R, and thus the third term actually equals
to one! Putting these together, we can get the famous entropy change
decomposition in time inhomogeneous constant-noise diffusion processes,
\begin{equation}
S_{\mathrm{tot}}=\ln\frac{p_{0}(\mathbf{X}_{0})}{p_{\tau}(\mathbf{X}_{\tau})}+Q=\Delta S+Q.\label{eq: Stot diffusion}
\end{equation}
With this, it is obvious that $S_{\mathrm{tot}}$ is additive in time.

We also note that the difference between $S_{\mathrm{T}}$ and $S_{\mathrm{tot}}$
in processes with time symmetric $\mathbf{b}$, \emph{i.e.} when $\mathbf{b}^{\mathrm{R}}=\mathbf{b}$,
is
\begin{equation}
S_{\mathrm{T}}-S_{\mathrm{tot}}=\ln\frac{p_{\tau}(\mathbf{X}_{\tau})}{p_{0}(\mathbf{X}_{\tau})}.\label{eq:ST-Stot  in diffusion}
\end{equation}
This is consistent with what we had in Equation (\ref{eq: dP/dPD in homogeneous}),
leading one to conclude $\mathbb{E}[S_{\mathrm{T}}]\ge\mathbb{E}[S_{\mathrm{tot}}]$
when $\mathbf{b}^{\mathrm{R}}=\mathbf{b}$.

\subsection{Housekeeping heat and Non-adiabatic Entropy Production}

Using the CPM perspective to get $Q_{\mathrm{hk}}$ has already been
rigorously studied in \cite{jiang_mathematical_2004}, here we shall
briefly revisit the derivation and rely on the relation $S_{\mathrm{tot}}=Q_{\mathrm{hk}}+S_{\mathrm{na}}$
to get the formula for $S_{\mathrm{na}}$.

In time inhomogeneous diffusion process, we could consider the instantaneous
stationary probability density $\pi_{t}$ such that 
\begin{equation}
\nabla\cdot\mathbf{J}[\pi_{t}]=0\label{eq: divergent free}
\end{equation}
where $\mathbf{J}[\pi_{t}]=\mathbf{b}(\mathbf{X}_{t},t)\pi_{t}-\mathbf{D}\nabla\pi_{t}$.
From the fact that the adjoint process is given by reversing $\mathbf{J}[\pi_{t}]$,
we see that the effect of the operator $\dagger$ on $\mathbf{b}$
is given by 
\begin{equation}
\mathbf{b}^{\dagger}=-\mathbf{b}-2\mathbf{D}\nabla\left(-\ln\pi_{t}\right).\label{eq: b dagger}
\end{equation}
Then, we use the decomposition 
\begin{equation}
\frac{\mathrm{d}\mathbb{P}\ }{\mathrm{d}\mathbb{P}^{\dagger}}=\frac{\mathrm{d}\mathbb{P}\ \ }{\mathrm{d}\mathbb{P}^{\mathrm{B}}}\frac{\mathrm{d}\mathbb{P}^{\mathrm{B}}\ }{\mathrm{d}\mathbb{P}^{\dagger\mathrm{B}}}\frac{\mathrm{d}\mathbb{P}^{\dagger\mathrm{B}}}{\mathrm{d}\mathbb{P}^{\dagger}\ }\label{eq: decomposition of dP/dP dagger}
\end{equation}
and note that the second term is 1 since neither $\dagger$ nor B
changes the initial distribution and B kills the drift no matter it
is $\mathbf{b}$ or $\mathbf{b}^{\dagger}$. Also, the third term
can be evaluated by Equation (\ref{eq:GSN Strat}) by substituting
$\mathbf{b}$ with $\mathbf{b}^{\dagger}$. Further using Equation
(\ref{eq: divergent free}) to simplify the expression one got from
above, one would arrive 
\begin{align}
Q_{\mathrm{hk}} & =\int_{0}^{\tau}\left[\mathbf{D}^{-1}\mathbf{b}+\nabla\left(-\ln\pi_{t}\right)\right]\circ\mathrm{d}\mathbf{X}_{t}\nonumber \\
 & =Q-Q_{\mathrm{ex}}\label{eq: Qhk}
\end{align}
 where $Q_{\mathrm{ex}}=-\int_{0}^{\tau}\nabla\left(-\ln\pi_{t}\right)\circ\mathrm{d}\mathbf{X}_{t}$
is the excess heat dissipation in diffusion. The non-adiabatic entropy
production $S_{\mathrm{na}}$ would then be given by 
\begin{align}
S_{\mathrm{na}} & =S_{\mathrm{tot}}-Q_{\mathrm{hk}}=W_{\mathrm{ex}}-\Delta F\nonumber \\
 & =\int_{0}^{\tau}\frac{\partial\left(-\ln\pi_{t}\right)}{\partial t}\mathrm{d}t-\ln\frac{p_{\tau}(X_{\tau})\pi_{0}(X_{0})}{\pi_{\tau}(X_{\tau})p_{0}(X_{0})}.\label{eq: Sna}
\end{align}

\section{Discussion\label{Discuss}}

In this paper, we characterize the difference between the statistical
properties of the original stochastic process and the one after reversal
by a change of probability measure, an analog to Schrödinger's picture
on Quantum Mechanics \citep{qian_ternary_2019}. A change in statistical
properties from a physical operation is represented by an operator
$\nu$ operating on a probability measure $\mathbb{P}$ in the probability
measure space $\mathcal{P}$. \textit{\emph{With our mathematically
more general and concise CPM formalism, we have presented a comprehensive
study of the properties of EPs including FRs. Sufficient conditions
for the FRs of the four EPs are summarized in Table \ref{tab:Summary-of-the}.
Importantly, a hierarchy of the generality for FRs in general stochastic
processes can be revealed from our work: both IFT and GFR are generally
true; rDFT and iDFT require odd parity symmetry with different $\hat{S}_{\nu}$
as stated in \citep{seifert_stochastic_2012}; and TFT further requires
the CPM operator to be realized by an involutive map on the trajectory
space $\Omega$. This hierarchical structure of the domain of validity
for FRs reveals relation between FRs such as TFT implies iDFT.}}

\textit{\emph{We further demonstrate how to obtain other properties
of EPs from their logarithm RND definitions such as their martingale
properties and distinguish the difference between dissipation function
introduced by Evans and Searles \cite{evans_fluctuation_2002} and
the total entropy production. The ``paradox'' that the two EPs have
the same entropy production rate but with non-negative difference
in expectation for finite time interval in time homogeneous processes
is resolved by noting the failure of time additivity for the dissipation
function. Stochastic integration expressions for the two EPs are also
derived in general time inhomogeneous constant-noise diffusion to
better see the contradistinction of their physical meaning and properties.}}

It is important to note that throughout this paper, we have assumed
the state variables $X_{n}$ to have even parity under the time reversal,
\emph{i.e.} they are position-like physical quantities. One extension
to our work is to consider variables that have odd parity under time
reversal such as velocity \citep{spinney_nonequilibrium_2012,ge_time_2014,li_stochastic_2019}.

The unit of EP $S_{\nu}$ is chosen to be in the natural unit of information
(nat) throughout the paper \citep{crooks_entropy_1999,cover_elements_2006}.
And temperature of the heat reservoir is assumed to be constant. When
considering diffusion processes in Section \ref{sec:Entropy-Productions-in Diffusion},
we have also restricted ourselves to constant-noise diffusion processes.
If the strength of noise $\mathbf{\Gamma}$ varies in spacetime, then
the use of Girsanov theorem to derive the formula for $S_{\mathrm{T}}$
and $S_{\mathrm{tot}}$ becomes more involved since $\frac{\mathrm{d}\mathbb{P}_{0}}{\mathrm{d}\mathbb{P}_{0}^{\nu}}$
where $\nu=\mathrm{T}$ or $\mathrm{RT}$ becomes less straightforward.
We note that Jiang, Qian, and Qian have used Girsanov theorem to derive
the housekeeping heat $Q_{\mathrm{hk}}$ for autonomous, non-constant
noise diffusion in \cite{jiang_mathematical_2004}. With this and
replying on the fact that the formula of $S_{\mathrm{na}}$ does not
depend on $\mathbf{\Gamma}$, we can use the relation $S_{\mathrm{tot}}=Q_{\mathrm{hk}}+S_{\mathrm{na}}$
to argue that Equation (\ref{eq: Stot diffusion}) still holds for
autonomous, non-constant noise diffusion processes. To obtain the
integral equation for $S_{\mathrm{T}}$, however, will need other
methods. One may need the path integral formalism to obtain the probability
``density'' of a diffusion path \cite{onsager_fluctuations_1953},
or make use of the Riemann geometry introduced by $\mathbf{\Gamma}$
and $\mathbf{D}$ to consider a constant-noise diffusion on a curved
space \cite{fujita_onsager-machlup_1982}.

The general theory we presented in Section \ref{sec:General-Theories}
is in fact a general result for \emph{fluctuating relative entropy}
and its statistical properties. It can also be applied to entropies
defined in information theory \citep{cover_elements_2006}. For example,
suppose we have a finite state space $\mathcal{X}\equiv\Omega$, \emph{i.e.}
there is a fundamental state random variable $X$ that labels every
$\omega\in\Omega$ with a unique real number, we can then choose $P_{X}^{\nu}$
to be the uniform distribution, \emph{i.e. }$\mathbb{P}^{\nu}$ as
the Lebesgue measure, to get an entropy corresponding to the maximum
entropy $\ln\Vert\Omega\Vert$ minuses the fluctuating Shannon entropy
of the system, 
\begin{equation}
H(\omega)=\ln\Vert\Omega\Vert-\left[-\ln P_{X}(X(\omega))\right]\label{eq:-39}
\end{equation}
where $\Vert\Omega\Vert$ represents the size of the sample space.
Another example would be to have $\Omega=\mathcal{X}\otimes\mathcal{Y}$
and consider the fluctuating mutual information between two random
variables $X$ and $Y$, 
\begin{equation}
I(\omega)=\ln\frac{P_{X,Y}(X(\omega),Y(\omega))}{P_{X}(X(\omega))P_{Y}(Y(\omega))}.\label{eq:-40}
\end{equation}
Our theory immediately implies that both $H(\omega)$ and $I(\omega)$
have non-negative expectation, and admit IFT and GFR.

The combination of information theory and stochastic thermodynamics
is a natural application and extension to the theory \citep{landauer_irreversibility_1961,bennett_thermodynamics_1982,still_thermodynamics_2012,parrondo_thermodynamics_2015}.
One example would be to consider random transition matrices for stochastic
driving protocol. With randomness in transition matrices, the second
law of thermodynamics is refined by incorporating the mutual information
between the past and present, and the mutual information between the
present and the future, giving us a thermodynamics of prediction \citep{still_thermodynamics_2012}.
In our change-of-measure formalism, we could extend $\left(\Omega,\mathcal{F}\right)$
to include all possible transition matrices. Such future work could
be conducted by considering the theories of random dynamical system
for Markov chains \citep{ye_stochastic_2016,x._-f._ye_stochastic_2019}.
\begin{acknowledgments}
The authors thank Tyler Chen, Yu-Chen Cheng, Ian Ford, Hao Ge, Liu
Hong, Chris Jarzynski, Geng Li, Matt Lorig, Udo Seifert, David Sivak,
Lowell Thompson, Zhanchun Tu, and Yue Wang for many helpful discussions.
We also thanks the anonymous referees for their feedback and suggestions.
\end{acknowledgments}

\section*{Appendix}

\subsection*{A. Modern Probability Theory and Radon-Nikodym Derivative\label{A}}

Here we briefly introduce several key concepts of modern measure-theoretic
probability theory pioneered by A. N. Kolmogorov \citep{kolmogorov_foundations_2018}.
For recent textbook introductions, see \citep{durrett_probability:_2010,grimmett_probability_2001}.
One of the most important concept in the theory is that one specifies
the sample space $\Omega$, the $\sigma$-algebra $\mathcal{F},$
and a probability (measure) $\mathbb{P}$ to start any probabilistic
discussion on stochastic processes.

An outcome $\omega$ is a particular result of a random trial. A sample
space $\Omega$ is the collection of all possible outcomes. We can
think of $\Omega$ as the state space of our system of interest. In
this paper, a ``state'' of the system of interest would be a trajectory
of a given finite time interval $t$ in a stochastic process. $\Omega$
then becomes the space of trajectories.

An event of interest $A$ is a subset of the sample space that we
seek the probability of. For example, in this paper, it can be the
set of periodic trajectories $\omega\in\Omega$, \emph{i.e.} we might
ask ``what is the probability of having a periodic trajectory''.
The collection of all event of interests of $\Omega$ is called $\sigma$-algebra
\footnote{It is called $\sigma$-algebra because it considers countable union
instead of union of a finite number of sets. If the latter case, it
is called an algebra instead of an $\sigma$-algebra.}, usually denoted as $\mathcal{F}$.

Note that when we are interested in an event, the complement of it,
$A^{c}$, \emph{i.e.} when $A$ does not happen, is also under our
interest. Moreover, if we are already interested in a bunch of events
$A_{1},A_{2}^{c},\dots$, then the countable union of them, \emph{i.e.}
the event when at least one of it happen, is also interesting to us.
Note that with complement and countable union, the countable intersections
are automatically included in $\mathcal{F}$.

As a result, a $\sigma$-algebra of $\Omega$ as a collection of all
events of interest must have the following three properties in its
very definition: (1) containing the empty set $\emptyset$ where nothing
happen, (2) being closed under countable union, and (3) being closed
under complement. One of the many reasons to introduce $\sigma$-algebra
$\mathcal{F}$ along with the sample space $\Omega$ is that since
we are interested in knowing the probability of every events of interest,
which is given by a probability measure $\mathbb{P}$, we need the
collection of all events of interest $\mathcal{F}$ to actually define
a $\mathbb{P}$.

A probability measure $\mathbb{P}$ measures the probability of an
event of interest $A\in\mathcal{F}$ by assigning the event a real
value, $\mathbb{P}\{A\}$, between 0 and 1. Since an event $A\in\mathcal{F}$
is a subset of $\Omega$, the probability of an event $A$ can also
be represented as an expectation of the indicator function of an event
$A$, $\mathbb{I}_{A}(\omega)$, \emph{i.e.} 
\begin{equation}
\mathbb{P}\{A\}=\mathbb{E}[\mathbb{I}_{A}(\omega)]\tag{A1}\label{eq: Expectation nottion for probability}
\end{equation}
where $\mathbb{I}_{A}(\omega)$ gives $1$ if $\omega\in A$ and $0$
otherwise, and $\mathbb{E}[\cdot]$ denotes the expectation. This
expectation expression of $\mathbb{P}\{A\}$ is convenient in many
calculations, especially when we consider a change of probability
measure as shown in Equation (\ref{eq: pdf of as a expectation})
and (\ref{eq: change of measure in explicit form}).

Physical observable as a random variable $X(\omega)$ is not just
a function from $\Omega$ to $\mathbb{R}$. Since we are interested
in the statistical properties of $X(\omega)$, we would like to use
the probability measure $\mathbb{P}$ defined on the $\mathcal{F}$
of $\Omega$ to get the probability of an event such as $\left\{ X(\omega)\in(a,b);a,b\in\mathbb{R}\right\} $
or for $X(\omega)$ in any countable union/complements of open intervals.
The collection of any set that can be formed from open intervals in
$\mathbb{R}$, complement of them, and/or countable union of them
is called the \emph{Borel} \emph{sets} of $\mathbb{R}$, denoted as
$\mathcal{B}(\mathbb{R})$. The requirement of being able to use a
probability measure $\mathbb{P}$ to get that statistical properties
of $X(\omega)$ is then the requirement of \emph{measurability}. That
is, for $\mathbb{P}\{X(\omega)\in B\}$ to make sense, we need that
$\forall B\in\mathcal{B}(\mathbb{R})$, $\left\{ X(\omega)\in B\right\} $
as a set of $\omega$ to be in $\mathcal{F}$. Hence, a random variable
is not a function of $\Omega$ but a \emph{measurable function $X(\omega):\Omega\mapsto\mathbb{R}$
}defined on $\left(\Omega,\mathcal{F}\right)$ so that $\left\{ X(\omega)\in B\right\} \in\mathcal{F}$,
$\forall B\in\mathcal{B}(\mathbb{R})$.

The very fact that $X$ is defined on a pair $\left(\Omega,\mathcal{F}\right)$,
without a specific $\mathbb{P}$, is a manifestation of the Schrödinger's
picture of changes in statistical properties. For two different stochastic
processes, the change in the statistical properties of a random variable
$X(\omega)$, the difference in its distribution $P_{X}$ and $P_{X}^{\nu}$,
is due to the change of probability measure $\mathbb{P}\rightarrow\mathbb{P}^{\nu}$
\citep{qian_ternary_2019}. Without $\mathbb{P}$, the pair $\left(\Omega,\mathcal{F}\right)$
is called \emph{measurable space. }With a $\mathbb{P}$ defined, the
triple $\left(\Omega,\mathcal{F},\mathbb{P}\right)$ is called the
\emph{probability space.} In this paper, a stochastic process is specified
by a probability space.

With a given measurable space $\left(\Omega,\mathcal{F}\right)$,
there are many $\mathbb{P}$ that can be considered. Radon-Nikodym
derivative (RND) is a special random variable that can be used to
change one probability measure $\mathbb{P}$ to another $\mathbb{P}^{\nu}$,
denoted as $\frac{\mathrm{d}\mathbb{P}^{\nu}}{\mathrm{d}\mathbb{P}\ }(\omega)$.
As explained in the main context, the probability of a event $A\in\mathcal{F}$
in the new probability measure is given by the expectation representation
shown in Equation (\ref{eq: Expectation nottion for probability}),
\begin{equation}
\mathbb{P}^{\nu}\{A\}=\mathbb{E}^{\nu}[\mathbb{I}_{A}(\omega)]=\mathbb{E}[\frac{\mathrm{d}\mathbb{P}^{\nu}}{\mathrm{d}\mathbb{P}\ }(\omega)\mathbb{I}_{A}(\omega)].\tag{A2}\label{eq:-41}
\end{equation}
For $\Omega$ whose $\omega$ in $\Omega$ can be 1-1 labeled by $x\in\mathbb{Z}$,
we have the RND reduced to the ratio of probability mass functions
$P^{\nu}(x)/P(x)$ since 
\begin{align*}
\mathbb{P}^{\nu}\{A\} & =\mathbb{E}^{\nu}[\mathbb{I}_{A}]=\sum_{x\in\mathbb{Z}}\mathbb{I}_{A}(x)P^{\nu}(x)\tag{A3a}\\
 & =\sum_{x\in\mathbb{Z}}\mathbb{I}_{A}(x)\frac{P^{\nu}(x)}{P(x)}P(x)\tag{A3b}\\
 & =\mathbb{E}[\mathbb{I}_{A}(x)\frac{P^{\nu}(x)}{P(x)}].\tag{A3c}
\end{align*}
For those $\Omega$ whose $\omega$ in $\Omega$ can be 1-1 labeled
by $x\in\mathbb{R}$ such as diffusion processes, we have the RND
reduced to the ratio of probability density functions $\rho^{\nu}(x)/\rho(x)$
since 
\begin{align}
\mathbb{P}^{\nu}\{A\} & =\mathbb{E}^{\nu}[\mathbb{I}_{A}]=\int_{\mathbb{R}}\mathbb{I}_{A}(x)\rho^{\nu}(x)\mathrm{d}x\tag{A4a}\nonumber \\
 & =\int_{\mathbb{R}}\mathbb{I}_{A}(x)\frac{\rho^{\nu}(x)}{\rho(x)}\rho(x)\mathrm{d}x\tag{A4b}\nonumber \\
 & =\mathbb{E}[\mathbb{I}_{A}(x)\frac{\rho^{\nu}(x)}{\rho(x)}].\tag{A4c}
\end{align}

\subsection*{B. Entropy Production Rates for $S_{\mathrm{T}}$ and $S_{\mathrm{tot}}$\label{B}}

With our trajectory-based definitions for dissipation function $S_{\mathrm{T}}$
and total entropy production $S_{\mathrm{tot}}$ given as 
\begin{equation}
S_{\mathrm{T}}(\omega)=\ln\frac{\mathrm{d}\mathbb{P}\ \thinspace}{\mathrm{d}\mathbb{P}^{\mathrm{T}}}(\omega)\text{ and }S_{\mathrm{tot}}(\omega)=\ln\frac{\mathrm{d}\mathbb{P}\ \ \ }{\mathrm{d}\mathbb{P}^{\mathrm{RT}}}(\omega)\tag{B1}\label{eq:-44}
\end{equation}
where $\omega=x_{0}x_{1}\cdots x_{t}$, here we show that although
the two EPs differ in finite time interval as shown in our main context,
they have the same rate in expectation in infinitesimal time interval
$t\rightarrow0$.

With an infinitesimal time interval $t\rightarrow0$, we only need
to consider one infinitesimal time step where the system state goes
from $i$ to $j$, 
\begin{equation}
S_{\mathrm{T}}(i,j)=\ln\frac{P_{0}(i)M(j|i)}{P_{0}(j)M(i|j)}\tag{B2}\label{eq:-45}
\end{equation}
and 
\begin{equation}
S_{\mathrm{tot}}(i,j)=\ln\frac{P_{0}(i)M(j|i)}{P_{t}(j)M(i|j)}\tag{B3}.\label{eq:-47}
\end{equation}
The difference between the two EPs is 
\begin{equation}
D(i,j)\coloneqq S_{\mathrm{T}}(i,j)-S_{\mathrm{tot}}(i,j)=\ln\frac{P_{t}(j)}{P_{0}(j)}\tag{B4}.\label{eq:-46}
\end{equation}
With $t\rightarrow0$, $M$ is approaching to an identity matrix,
\begin{equation}
M(j|i)=\delta_{i,j}+q(j|i)\cdot t+O(t^{2}).\tag{B5}\label{eq:-48}
\end{equation}
Up to the linear order, we can write 
\begin{align}
P_{t}(j) & =\sum_{k}P_{0}(k)M(j|k)\tag{B6a}\nonumber \\
 & =P_{0}(j)+t\sum_{k}P_{0}(k)q(j|k)+O(t^{2}).\tag{B6b}\label{eq:-49}
\end{align}

In the literature, entropy production rate of $S_{\mathrm{T}}$ and
$S_{\mathrm{tot}}$ is defined as their limiting re-scaled expectation:
$\lim_{t\rightarrow0}\frac{\mathbb{E}[S_{\mathrm{T}}]}{t}$ and $\lim_{t\rightarrow0}\frac{\mathbb{E}[S_{\mathrm{tot}}]}{t}$.
Therefore, let us compute the expectation of $S_{\mathrm{T}}$, $S_{\mathrm{tot}}$,
and $D$ as an asymptotic series of small $t$. Since $D=S_{\mathrm{T}}-S_{\mathrm{tot}}$,
we have 
\begin{equation}
\mathbb{E}[D]=\mathbb{E}[S_{\mathrm{T}}]-\mathbb{E}[S_{\mathrm{tot}}]\tag{B7}\label{eq:-50}
\end{equation}
and we will compute $\mathbb{E}[D]$ and $\mathbb{E}[S_{\mathrm{T}}]$.

For $\mathbb{E}[S_{\mathrm{T}}]$, we have 
\begin{align}
\mathbb{E}[S_{\mathrm{T}}] & =\sum_{i,j}P_{0}(i)M(j|i)\ln\frac{P_{0}(i)M(j|i)}{P_{0}(j)M(j|i)}\tag{B8a}\label{eq:-84}\\
 & =0+t\sum_{i,j;i\neq j}P_{0}(i)q(j|i)\ln\frac{P_{0}(i)q(j|i)}{P_{0}(j)q(j|i)}\tag{B8b}\label{eq:-51}
\end{align}
And for $\mathbb{E}[D]$, we have 
\begin{align}
\mathbb{E}[D] & =\sum_{i,j}P_{0}(i)M(j|i)\ln\frac{P_{t}(j)}{P_{0}(j)}\tag{B9a}\nonumber \\
 & =\sum_{j}P_{t}(j)\ln\frac{P_{t}(j)}{P_{0}(j)}.\tag{B9b}
\end{align}
With $P_{t}(j)=P_{0}(j)+t\sum_{k}P_{0}(k)q(j|k)+O(t^{2})$, we compute
\begin{align}
\ln\frac{P_{t}(j)}{P_{0}(j)} & =\ln\frac{P_{0}(j)+t\sum_{k}P_{0}(k)q(j|k)+O(t^{2})}{P_{0}(j)}\tag{B10a}\nonumber \\
 & =t\frac{\sum_{k}P_{0}(k)q(j|k)}{P_{0}(j)}+O(t^{2}).\tag{B10b}\label{eq:-52}
\end{align}
We thus see that 
\begin{align}
\mathbb{E}[D] & =t\sum_{j}P_{t}(j)\frac{\sum_{k}P_{0}(k)q(j|k)}{P_{0}(j)}+O(t^{2})\tag{B11a}\nonumber \\
 & =t\sum_{k}P_{0}(k)\sum_{j}q(j|k)+O(t^{2})\tag{B11b}\nonumber \\
 & =0+O(t^{2}).\tag{B11c}\label{eq:-53}
\end{align}
Hence, the two EPs have the same entropy production rate 
\begin{align}
\mathrm{e_{p}} & =\lim_{t\rightarrow0}\frac{\mathbb{E}[S_{\mathrm{T}}]}{t}\tag{B12a}\nonumber \\
 & =\lim_{t\rightarrow0}\frac{\mathbb{E}[S_{\mathrm{tot}}]+\mathbb{E}[D]}{t}=\lim_{t\rightarrow0}\frac{\mathbb{E}[S_{\mathrm{tot}}]}{t}.\tag{B12b}\label{eq:-54}
\end{align}
in the infinitesimal time interval limit.

\bibliography{yangepPREresub.bbl}

\end{document}